\documentclass[article,12pt]{article}

\usepackage{amsmath,aas_macros}
\usepackage[utf8]{inputenc}
\usepackage{fullpage}
\usepackage{natbib}
\usepackage{graphicx,inputenc,multicol,subfig,longtable,palatino,avant}
\usepackage{hyperref}
\usepackage{natbib}

\usepackage{graphicx}%
\usepackage{multirow}%
\usepackage{amsmath,amssymb,amsfonts}%
\usepackage{amsthm}%
\usepackage{mathrsfs}%
\usepackage[title]{appendix}%
\usepackage{xcolor}%
\usepackage{textcomp}%
\usepackage{manyfoot}%
\usepackage{booktabs}%
\usepackage{listings}%
\usepackage{pdflscape}	
\usepackage{hyperref}

\date{}

\title{Characterizing proton polytropic indices inside near-Earth magnetic clouds and ICME sheaths}
\author{Debesh Bhattacharjee\textsuperscript{1*}, Saikat Majumder\textsuperscript{2} \& Prasad Subramanian\textsuperscript{3} \\ \textsuperscript{1} \small {School of Physics and Astronomy, Kelvin Building, University of Glasgow, Glasgow, G12 8QQ, UK } \\ 
\textsuperscript{2} \small{Digantara Research and Technologies Pvt. Ltd, Hebbal Kempapura, Bengaluru, Karnataka-560024, India} \\ 
\textsuperscript{3} \small{Department of Physics, Indian Institute of Science Education and Research Pune,} \\ \small{Dr. Homi Bhabha Road, Pashan, Pune-411008, India} \\
\textsuperscript{*} \small{\textit{debesh.bhattacharjee@glasgow.ac.uk}}}

\begin{document}
	
	\maketitle

	\begin{abstract}
		The thermodynamics of interplanetary coronal mass ejections (ICMEs) is often described using a polytropic process. Estimating the polytopic index ($\gamma$)  allows us to quantify the expansion or compression of the ICME plasma arising from changes in the plasma temperature. In this study, we estimate $\gamma$ for protons inside the magnetic clouds (MCs), their associated sheaths, and ambient solar wind for a large sample of well-observed events observed by the Wind spacecraft at 1 AU. We find that $\gamma$ shows a high ($\approx 1.6$) - low ($\approx 1.05$) - high ($\approx 1.2$) behavior inside the ambient solar wind, sheath, and MCs, respectively. We also find that the proton polytropic index is independent of small-scale density fluctuations. Furthermore, our results show that the stored energy inside MC plasma is not expended in expanding its cross-section at 1 AU. The sub-adiabatic nature of MC plasma implies external heating - possibly due to thermal conduction from the corona. We find that the heating gradient per unit mass from the corona to the protons of MC at 1 AU is $\approx 0.21$ erg cm$^{-1}$ g$^{-1}$ which is in agreement with the required proton heating budget.
		

	\end{abstract}
	
	Interplanetary Coronal Mass Ejections --- Sheaths --- Polytropic Process --- Proton Heating.
	

	 \section{Introduction}
	 \label{S - Intro}
	 
	 Solar coronal mass ejections (CMEs) directed towards the Earth and their associated sheath regions are some of the primary drivers of space weather disturbances \citep{1988tsurutani,2011guo,2019kilpua}. CMEs during their propagation phase, are often known as interplanetary CMEs or ICMEs. A typical ICME comprises of two major parts: i) sheath - a compressed turbulent high plasma $\beta$ solar wind structure in front of magnetic clouds \citep{2017kilpua,2023Ala-Lahti} and ii) magnetic clouds (MCs): a magnetic flux rope structure associated with strengthened magnetic ﬁeld, smooth magnetic rotation, and suppressed proton temperature \citep{1981burlaga}.  The enhanced turbulence inside the sheath plasma can significantly intensify the intensity of geomagnetic storms\citep{2003borovsky}. It is therefore of no surprise that predictions of arrival times and speeds of ICMEs and sheaths at the Earth form one of the most active and interesting fields of research in solar astrophysics \citep{2001Gopal,2018Riley,2021alahati}. The energy stored in ICME magnetic fields acts as the primary reservoir of energy \citep{2010Mandrini} and a part of it is thought to be expended in cooling the plasma during its expansion through the heliosphere. Characterizing ICMEs and ICME-driven sheaths and understanding their internal thermodynamic processes close to the Earth are, therefore, one of the key ways to advance space weather research.\\
	 Extreme ultraviolet (EUV) observations of CMEs close to the Sun by the Solar and Heliospheric Observatory (SOHO) spacecraft show that the CME temperature is higher than the ambient solar wind temperature, implying that the inner corona is transferring the energy to the CME plasma \citep{2006kohl,2010bemporad}. On the other hand, in-situ observations using different spacecrafts (e.g., WIND, Helios, Parker Solar Probe (PSP), etc.) provide us with an unique opportunity to understand the various internal thermodynamic processes and plasma turbulence of ICMEs and their associated sheaths \citep{2020mishra,2023Debesh,2022debesh, 2023Soumyaranjan,2022dayeh,2024dhamane}.\\
	 Estimation of polytropic index ($\gamma$) using the density and temperature profiles of ICME plasma has been one of the key ways to understand their thermodynamics \citep{1993Osherovich,1999gosling, 2001riley,2022dayeh}. $\gamma = 1$ suggests that the ICME plasma is undergoing an isothermal transition and the plasma is connected to the external heat reservoir (i.e. the solar corona). 
	 On the other hand, $\gamma = 5/3$, would imply that the ICME is expanding adabatically and that there is no heat exchange with the surrounding. $\gamma < 5/3$ accounts for the presence of an unspecified heating mechanism in the plasma, such as an external heating source and / or a local heating due to plasma turbulence. \cite{1993Osherovich} found that the proton polytropic index inside ICMEs is $\approx 1.2$ and the electron polytropic index is $\approx 0.48$. \cite{2022dayeh} used density-temperature scatterplots for a set of 336 events to determine the polytropic index inside ICMEs, sheath, and the pre and post-event regions. They found that inside the MCs and sheaths, the polytropic index is $\approx 1.54$ and $\approx 1.40$, respectively. Recently, \cite{2025debesh} have studied the proton polytropic index inside 27 most geo-effective ($D_{\rm st} < -75$ nT) near-Earth ICMEs of solar cycle 24 by analyzing their turbulent energy cascade. They have found the polytropic index to be $\approx 1.35$. \cite{2009Wang}, \cite{2018Wageesh}, and \cite{2023Soumyaranjan} have used the observed self-similar expansion of CMEs along with an assumption of self-similarity for the proton polytropic index to surmise that it varies from $\approx 5/3$ near the Sun to $\approx 1.2$ near the Earth. Recent studies like \cite{2025katsavrias} investigate 401 ICMEs between 1995 and 2001 using the in-situ data from the Wind spacecraft and find that the proton polytropic index of the ICME plasma depends on the magnetic orientation of flux rope. Their results show that flux ropes with rotation above 90 degrees exhibit sub-adiabatic behavior.\\ 
	 \cite{2018livadiotis} has found that $\gamma$ of solar wind remains independent of the bulk solar wind speed. The relation between turbulent fluctuations and solar wind $\gamma$ has been studied by \cite{2019Livadiotis}. They  estimated the radial profiles of solar wind heating in the inner and outer heliosphere. \cite{2020Nicolau} used the PSP data and showed that small-scale turbulent fluctuations of solar wind density between 0.17 AU and 0.8 AU follow a polytropic nature with $\gamma \approx 2.7$.  However, to the best of our knowledge, no studies have been performed showing the connection between turbulent density fluctuations and proton $\gamma$ inside the MC and sheath plasma for a large sample of near-Earth events. In this paper, we are leveraging the in-situ data from a large set of well-observed events to compare the proton polytropic indices inside the MC, sheath, and ambient solar wind plasma at 1 AU. Using the estimated values of $\gamma$ inside the MC and sheath, we investigate whether there is any relation between the polytropic process of MC, sheath plasma and small-scale density turbulence. We further examine whether the energy stored in the MC plasma is expended in its cross-sectional expansion at 1 AU.
	 Finally, we quantify the heating gradient for the protons of the MC plasma at 1 AU to understand how much energy the protons are getting from the corona. The outcomes of this study will help us better understand the thermodynamics of ICME and sheath plasma at 1 AU, thereby aiding us to better characterize their impacts on space weather.\\
	 The rest of the paper is organized as follows: $\S~\ref{S - data}$ describes in detail the datasets used for the analysis. $\S~\ref{S - results}$ has four subsections. $\S~\ref{S - polytropic index}$ shows polytropic index estimation process, while $\S~\ref{S - density modulation index}$ describes its relationship with turbulent density fluctuations. The relation between the polytropic index and the cross-sectional speed of the MC is described in $\S~\ref{S - vexp}$. We estimate the radial heating gradient in $\S~\ref{S - heating gradient}$. Finally, $\S~\ref{S - conclusion}$ contains the conclusion and discussion.
	  \begin{figure}
	 	
	 	\includegraphics[width=0.90\textwidth , height = 0.42\textwidth , scale=0.9]{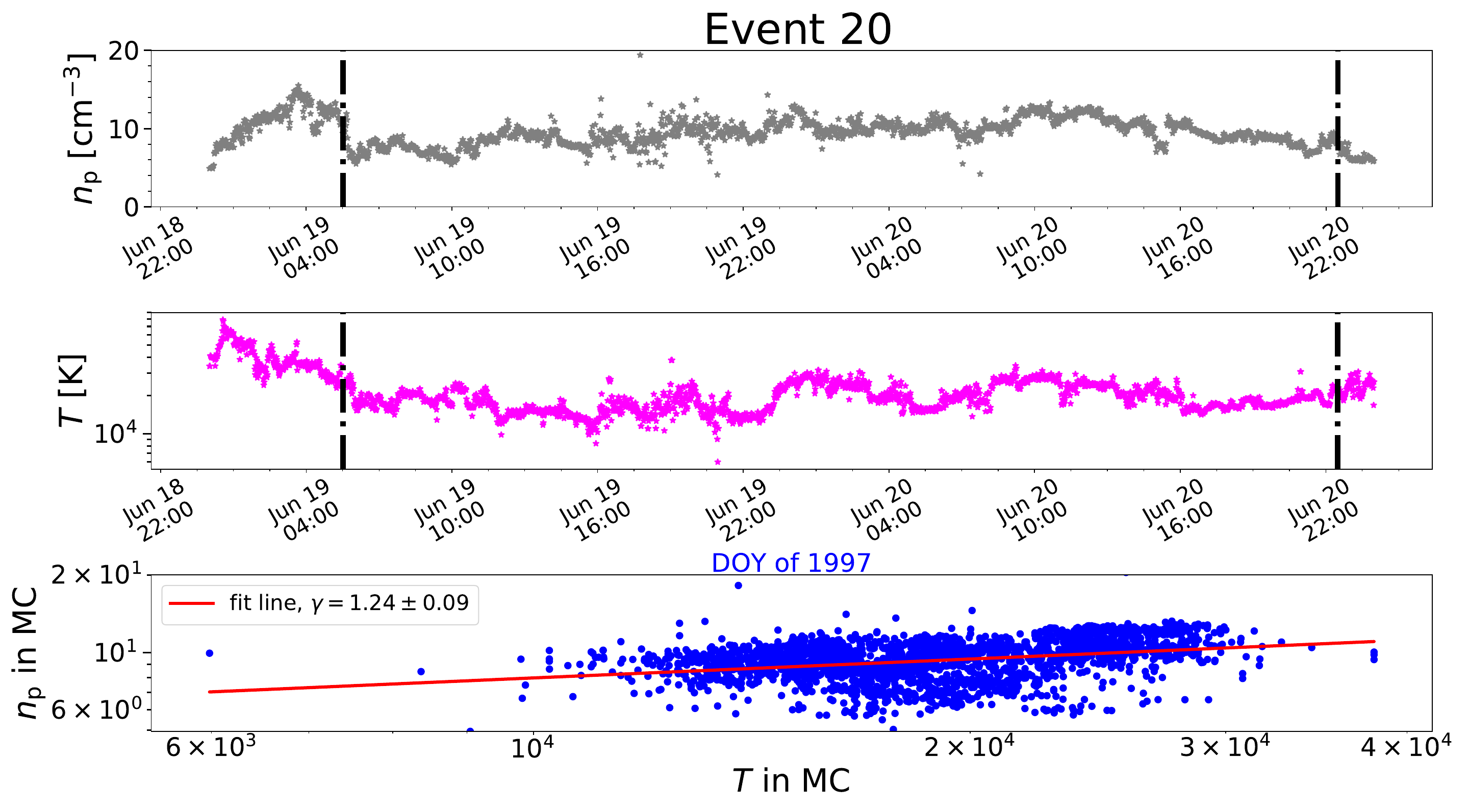}

	 	\caption{An example (event 20 in Table~\ref{S - Table A}) showing the estimation of proton polytropic index ($\gamma$) inside MC plasma by fitting the time profiles of proton number density ($n_{\rm p}$) and temperature ($T$) using Eq~\ref{eq poly2}. The top and the middle panels of this figure show the time profiles of $n_{\rm p}$ and $T$, respectively. The dotted vertical lines mark the start and end of MC for this event. The bottom plot shows the scatter plot between $n_{\rm p}$ and $T$. The red fitted straight line between these two parameters gives us the estimate of $\gamma$. For this particular event, $\gamma = 1.24 \pm 0.09$.} \label{Fig: event example}
	 \end{figure}
	 \section{Data}
	 \label{S - data}
	 We use the sample of 152 well-observed ICMEs used in \cite{2023Debesh, 2023Debesh2} and observed by the Wind spacecraft (\url{https://wind.nasa.gov/}) close to the Earth ($\approx 1$AU). First we have selected all the events observed between 1995 and 2015 that are listed as Fr or Fr+ in \cite{2018NC} and the Wind ICME catalog (\url{https://wind.nasa.gov/ICMEindex.php}). These events are the magnetic clouds (MCs) presumably conform best to the expectations of a flux rope structure \citep{2016NC,2018NC}. Fr events imply MCs with a single magnetic field rotation between 90$^\circ$ and 180$^\circ$ , while F+ events represent MCs with a single magnetic field rotation greater than 180$^\circ$ . We have further shortlisted this list considering events which are neither preceded nor followed by any other ICMEs or ejecta within a window of two days ahead of and after the event under consideration. This provides us a way to exclude possibly interacting events from the chosen data set. For completeness, we have listed the events in Table~\ref{S - Table A}. We use the nomenclature of \cite{2019NCSoPh}, where the location marked ``ICME start'' denotes the start of the sheath and the part between ``MC start'' and ``MC end'' is taken to be the body of the CME. For this study, we consider two different solar wind backgrounds (BGs) based on the selection criteria of \cite{2023Debesh2}. The first background (denoted as BG1 in this paper) is taken to be a 24-hour stretch of quiet solar wind in the five days preceding each event. BG1 has the following criteria: (i) the root-mean-square (rms) fluctuations of the plasma velocity for this 24-hour period should not exceed 10\%
	 of the mean value, (ii) the rms fluctuations of the total magnetic field
	 for this 24-hour period must not exceed 20\% of the mean
	 value, and (iii) there should not be no magnetic field rotations. Criteria (i) and (ii) ensure that the chosen background is quiet. Criterion (iii) distinguishes the background from the MCs, because MCs are characterized by
	 large rotations of magnetic field components and low-plasma beta. On the other hand, second background (denoted as BG2) is chosen to be the 24-hour solar wind stretch immediately before the ``ICME" start. For this study, we use the 1-minute cadence data from the 
	 Solar Wind Experiment (SWE) \citep{1995swe} instrument onboard the Wind spacecraft.
	 The time profile of plasma number density inside the MC, sheath and solar wind backgrounds for each of these events is provided by the Wind/SWE instrument. We use the time series of the proton temperature from the OMNI database (\url{https://omniweb.gsfc.nasa.gov/}) considering the 1-minute cadence data associated with the Wind plasma Key Parameters (KP). This is because the data from the SWE instrument (\url{https://wind.nasa.gov/mfi_swe_plot.php}) do not provide us with the time profile of proton temperature. We note that one can also derive temperature indirectly from the proton thermal speed ($v_{\rm th}$) measured by the Wind/SWE. We have checked that the temperature profiles for all of the studied events obtained from OMNI are inline with the observation from the SWE instrument.
	 We do not have OMNI temperature data for two events (event numbers 63 and 73) out of the 152 events enlisted in Table~\ref{S - Table A}. Therefore, our MC list consists of 150 events. Out of these 150 events, for 22 events, the `ICME start' coincides with `MC' start, implying no sheath region. Therefore, we only consider ($150-28 =$) 124 events for the analysis of the sheath. In summary, we use 150 MCs and BGs, and 124 sheaths for the current analysis. We compute the magnitudes of turbulent density modulation index (denoted as $\delta n/ n$) inside the ICMEs and sheaths following the method adopted by \cite{2023Debesh}. In the definition of the density modulation index, the quantity $n$ refers to the average value of the plasma number density in the moving box ($t_{\rm box}$), while $\delta n$ refers to the root-mean-square (rms) value of the fluctuations in the plasma number density in the box. 
	 
	 
	 \section{Results}
	 \label{S - results}

	 \begin{figure}
	 	
	 	\includegraphics[width=0.9\textwidth , scale=0.9]{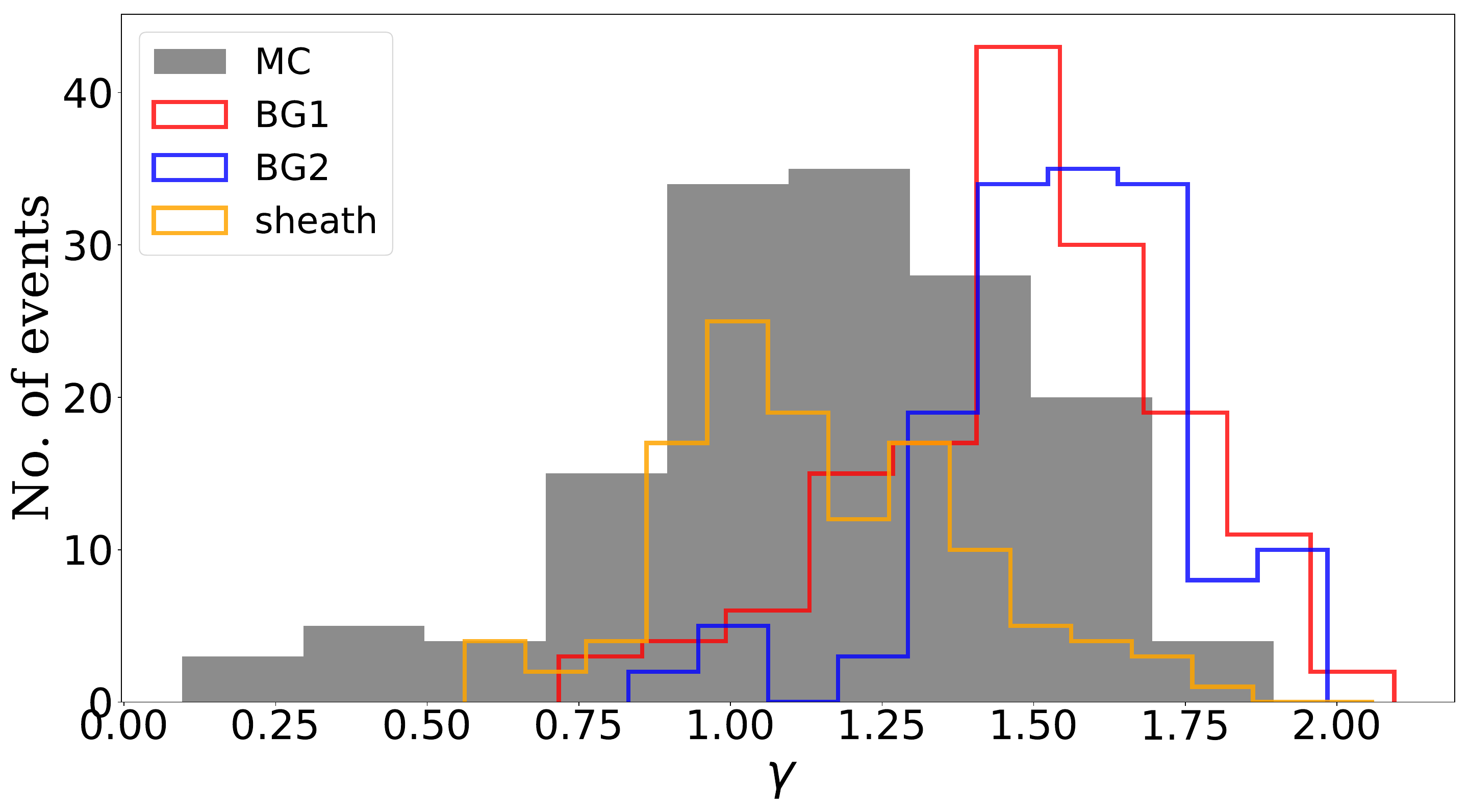}

	 	\caption{Histograms showing the distribution of polytropic indices inside MCs -- solid grey histogram, sheaths -- hollow yellow histogram, and two ambient solar wind backgrounds (BG1 and BG2) -- red and blue hollow histograms, respectively.} 
	 \label{Fig: polyhisto}
	 \end{figure}

	 This section briefly describes the results of the current study in addressing the items mentioned in \S~\ref{S - Intro}.
	 
	 \subsection{Estimation of polytropic index}
	 \label{S - polytropic index}
	 We assume that the number density ($n_{\rm p}$), temperature ($T$), pressure ($p$) and volume ($V$) obey
	 \begin{equation}
	 \label{eq poly1}
	 pV^{\gamma} = C  \,\, {\rm or }  \\
	 T \propto n_{\rm p}^{\gamma -1},
	 \end{equation}
	 
	 where $\gamma$ represents the polytropic index, $C$ is a constant and we have used the ideal gas law $PV = n_{p} k_{B} T$. In a polytropic process, the ratio between the energy supplied or transferred to the system as heat and the work done by the system remains constant \citep{1963Parker,1967Chandrasekhar}. In a certain condition, when there is no exchange of heat with the environment, the polytropic transition of the system is known as an adiabatic process. For an adabatic process, $\gamma = \gamma_a = c_p / c_v$, where $c_p$ and $c_v$ denotes specific heat under constant pressure and specific heat under constant volume, respectively. In such a case, one can write $\gamma_a = 1 + 2/d_{\rm f}$, where $d_{\rm f}$ represents the degrees of freedom of a particle. Therefore, for a plasma particle with three degrees of freedom undergoing an adiabatic transition, $\gamma = \gamma_a = 5/3$. However, in the presence of heating, the thermodynamic process is not adiabatic and therefore, $\gamma \neq \gamma_a$. Taking the natural logarithm in both sides of Eq~\ref{eq poly1}, one can write
	 \begin{equation}
	 \label{eq poly2}
	 ln(T) = (\gamma -1)ln(n_{\rm p}) + C_1 
	 \end{equation}
	 where $C_1$ is a constant. A polytropic relationship therefore provides us with insight into the heating or cooling of the plasma system without solving the
	 energy equation, which can often be complicated \citep{2006kartalev}.

\begin{figure}
	 
	 \includegraphics[width=0.9\textwidth , scale=0.9]{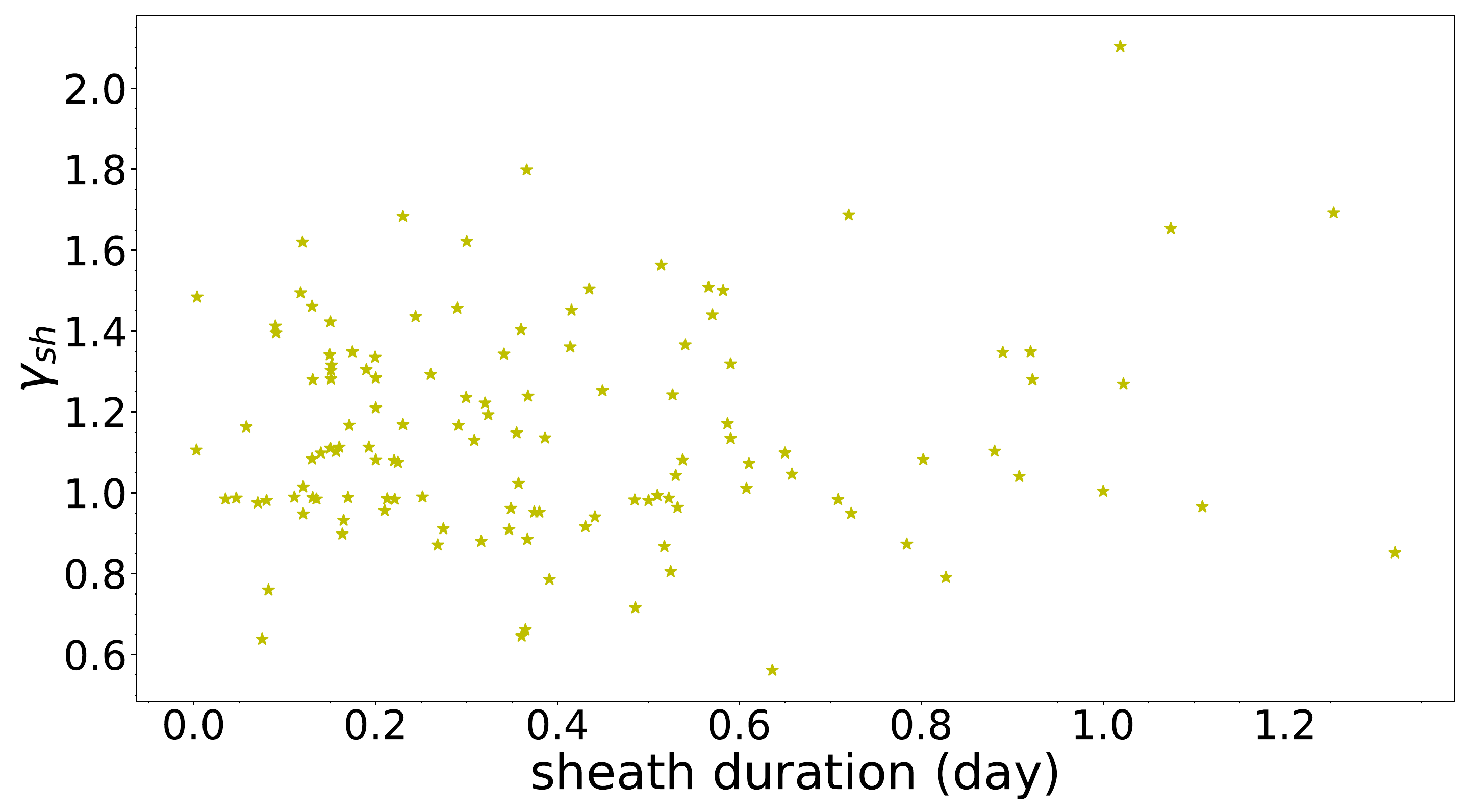}

	 \caption{Figure showing the plot between sheath duration (day) and polytropic index inside sheath ($\gamma_{\rm sh}$) for all the chosen sheaths in this study. This plot shows that there is no correlation between these two parameters, suggesting the $\gamma_{\rm sh}$ associated with near-Earth ICMEs remains independent of the sheath thickness.} 
	 \label{Fig: sh_duration}
 \end{figure}

	 We compute $\gamma$ for protons inside the MCs, sheaths and ambient solar wind backgrounds using the fit between the time profiles of proton number density, $n_p$ and temperature, $T$ at the location of the Wind spacecraft. An example of such fit is shown in the three subplots of Fig~\ref{Fig: event example}. These plots show how $\gamma$ is computed inside the MC plasma for event 20 of Table~\ref{S - Table A}. The values of $\gamma$ with their error estimates for 150 MCs, 150 backgrounds (BG1 and BG2), and 124 sheaths are listed in Table~\ref{S - Table B}. 
	 The results of estimated $\gamma$ inside MCs, sheaths and two ambient solar wind bckgrounds (BG1 and BG2) for all of our studied events enlisted in Table~\ref{S - Table A} are shown in the four histograms of Fig~\ref{Fig: polyhisto}. The mean, median, and most probable value (mpv) of $\gamma$ for 150 MCs are 1.16, 1.20, and 1.20, respectively. In this paper, we have used the `auto' option in matplotlib to determine the optimum bin size for each histogram of this analysis \citep{2025deep}. The mean, median, and mpv of $\gamma$ for 124 sheaths are 1.15, 1.10, and 1.05, respectively. We find that the mean, median, and the mpv for 150 BG1 are 1.49, 1.52, and 1.53, respectively. For BG2, the values are 1.56, 1.59, and 1.63, respectively. These findings suggest that the protons inside MCs and sheaths are close to isothermal compared to the protons in ambient solar wind plasma. However, we note that the $\gamma$ of MC has a larger distribution than that of the sheaths and solar wind backgrounds. Considering the mpvs, we find that the pre-ICME solar wind (BG2) has the $\gamma = 1.63$ which is close to the adiabatic value ($\gamma_a = 1.67$). Then as we enter in the sheath region, mpv of $\gamma$ drops to 1.05 which is close to the isothermal value ($\gamma = 1$) and during the passing of the MC, the mpv of $\gamma$ rises to 1.20. This high-low-high feature of $\gamma$ is in agreement with the outcomes of \cite{2022dayeh}. However, we note that our estimated magnitudes of mpvs of $\gamma$ are lower than their estimates. This could be because our selection criteria for the events are different from theirs. From the Fig~\ref{Fig: sh_duration}, we note that the polytropic index inside the sheath ($\gamma_{\rm sh}$) is not correlated with the sheath duration (Pearson's correlation coefficient ($r$) = 0.08 with p-value 0.3). A p-value $\leq 0.05$ suggests a 95\% confidence in the estimated value of $r$ \citep{pvalue}. This suggests that the near-isothermal behavior of sheath plasma at 1 AU is not associated with the sheath thickness. Our estimated proton polytropic index in MC plasma (the mpv value, $\gamma = 1.20$) is in agreement with the findings of proton polytropic index ($\gamma_p \approx 1.2$) inside MCs at 1 AU by \cite{1993Osherovich}. The fact that the protons of MC plasma behaves isothermally at 1 AU is also in agreement with the predictions of FRIS model \citep{2024khuntia2}. Though both the MC and sheath plasma show a sub-adiabatic behavior ($\gamma < 5/3$) at 1 AU, our findings suggest that the protons in sheath plasma are more isothermal than that of the MC plasma.
	 
\begin{figure}[h!]

\includegraphics[width=0.5\textwidth , scale=0.9]{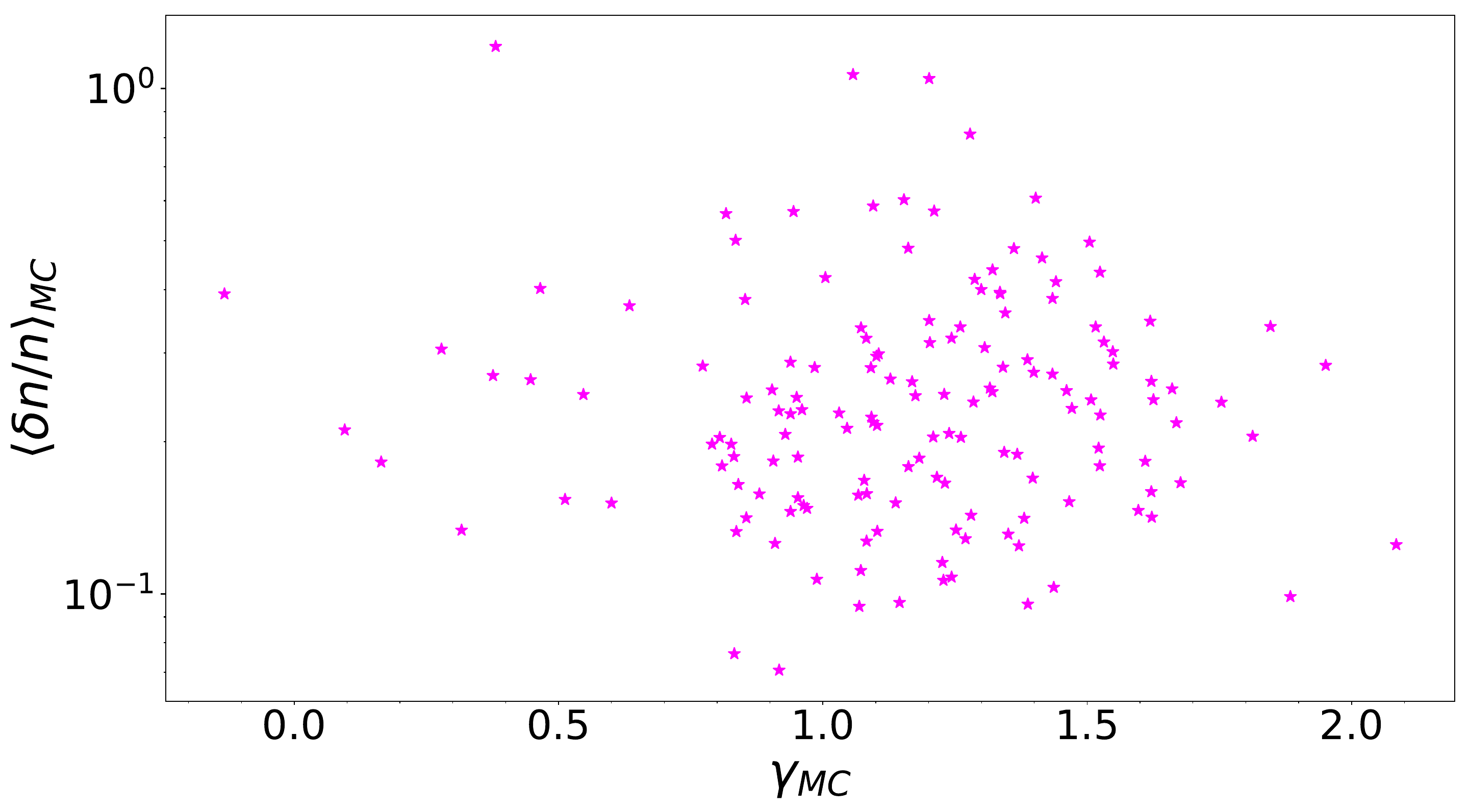}
\includegraphics[width=0.5\textwidth , scale=0.9]{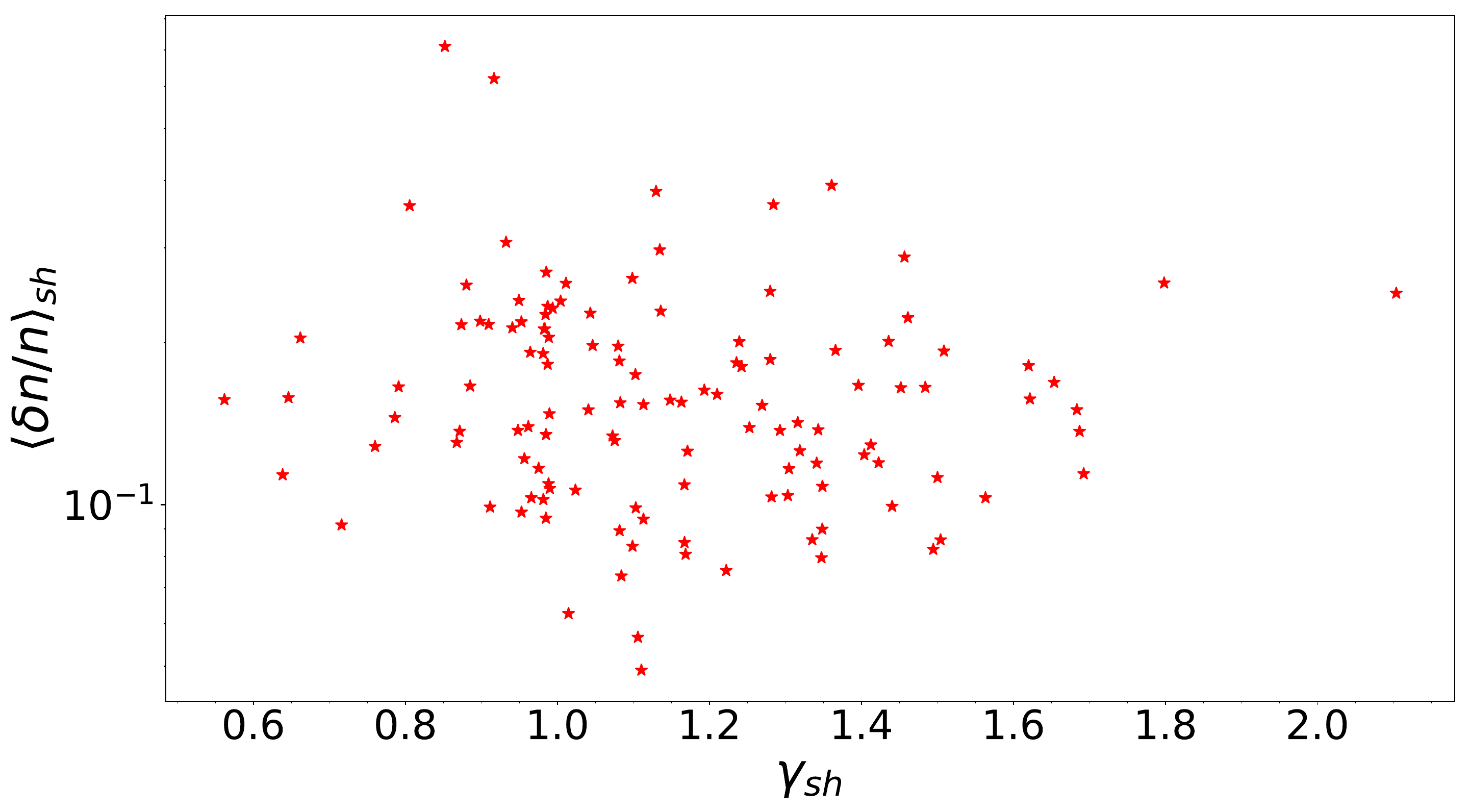}

\caption{The top panel shows the plot between the temporal average of turbulent density modulation index inside MC ($\langle \delta n / n\rangle_{\rm MC}$) and the proton polytropic index of MC plasma ($\gamma_{\rm MC}$). The bottom panel shows the plot between the tempral average of turbulent density modulation index inside sheath ($\langle \delta n / n\rangle_{\rm sh}$) and the proton polytropic index of sheath plasma ($\gamma_{\rm sh}$). Both the plots show that the density turbulence inside near-Earth MCs and sheaths remains independent of the value of proton polytropic index inside them.} 
\label{Fig: flucts}
\end{figure}

\subsection{Density modulation index}
\label{S - density modulation index}
The in-situ data from the Wind spacecraft provides us with the time profile of proton number density ($n_{\rm p}$) along the intercept in the spacecraft frame of reference. As a polytropic transition is associated with a macroscopic change in proton density, it is interesting to ask how $\gamma$ is affected by turbulent proton density fluctuations. To do that, we examine the behavior of $\delta n / n$ inside the MC and sheath plasma for the events listed in Table~\ref{S - Table A}. $\delta n / n$ quantifies the extent to which the proton density is modulated due to turbulent fluctuations, therefore, signifies the strength of the proton density turbulence \citep{2023Debesh}. We compute $\delta n / n$ inside MC by estimating the running mean and rms fluctuations of the proton density ($\delta n$) using $t_{\rm box} = 60$ minutes (hereafter abbreviated as mins) \citep{2023Debesh}. The duration of a typical sheath is shorter than that of a MC in our data set. For the estimation of $\delta n / n$ inside the sheath, we choose $t_{\rm box} = 20$ mins. This ensures that we are in the inertial regime of the turbulent spectrum with a sufficient number of data points to perform a statistical study. $t_{\rm box} = 60$ mins yields $\approx 25$ and $t_{\rm box} = 20$ mins yields $\approx 28$ temporal boxes inside MC and sheath, respectively. 

The temporal average of turbulent density modulation indices inside the MC ($\langle \delta n / n\rangle_{\rm MC}$), and sheath ($\langle \delta n / n\rangle_{\rm sh}$) are calculated using Eq 2 of \cite{2023Debesh}.
The results are shown in the top and bottom panels of Fig~\ref{Fig: flucts}. We find that the Pearson's correlation coefficient, $r$, between $\langle \delta n / n\rangle_{\rm MC}$ and $\gamma_{\rm MC}$ is -0.07 with p-value 0.4. Between $\langle \delta n / n\rangle_{\rm sh}$ and $\gamma_{\rm sh}$, we find $r = -0.11$ with p-value 0.2. Both of these findings show a very poor correlation between the turbulent density modulation index and the respective proton polytropic index. Therefore, these results suggest that the macroscopic thermodynamic changes (e.g, changes in $n_{\rm p}$, $p$, etc) in the protons of MC and sheath plasma at 1 AU are independent of the small-scale proton density fluctuations. 

\subsection{MC cross-sectional expansion speed}
\label{S - vexp}
A polytropic process involves a change in the volume with the change in the pressure via a certain way (see Eq~\ref{eq poly1}). Therefore, it is interesting to investigate how the polytropic index of the MC plasma ($\gamma_{\rm MC}$) correlates with the cross-sectional expansion speed ($v_{\rm EXP}$) of the MC at 1 AU. The time profile of plasma speed ($v$) obtained from the in-situ data provides us the opportunity to estimate $v_{\rm EXP}$ along the line of intercept between the MC and the Wind spacecraft. $v_{\rm EXP}$ is defined as \citep{2018NC,2022debesh}
\begin{equation}
	\label{eq vexp}
	v_{\rm EXP} = \frac{(v_{\rm s} - v_{\rm e})}{2}
\end{equation}
where $v_{\rm s}$ and $v_{\rm e}$ represent the start and end speeds of the MC boundary, as obtained from a linear fit of the time profile of plasma speed (see Fig 1 of \cite{2022debesh} for the reference). A high $v_{\rm EXP}$ would suggest that the front end of the MC is moving faster than the rear end, causing the cross-sectional area of the MC to expand. In other words, the volume of MC expands as it propagates through the heliosphere. A higher magnitude of $v_{\rm EXP}$ implies faster expansion. 
In \S~\ref{S - polytropic index} we find that protons in near-Earth MC plasma behave close to isothermal with $\gamma \approx 1.20$ (considering mpv). This suggests that MC plasma requires positive heating ($dQ >0$, $dQ$ is the measure of heat input to the system using the first law of thermodynamics) close to the Earth. The excess energy can be used to maintain the plasma temperature. This energy can also be expended in ICME expansion via energy absorption in the sub-adiabatic plasma. This motivates us to investigate whether a lower proton $\gamma$ is associated with a higher $v_{\rm EXP}$. To do that, we estimate the Pearson's correlation coefficient ($r$) between them for the MCs listed in Table~\ref{S - Table A}. 
\begin{figure}
	
	\includegraphics[width=0.90\textwidth , scale=0.9]{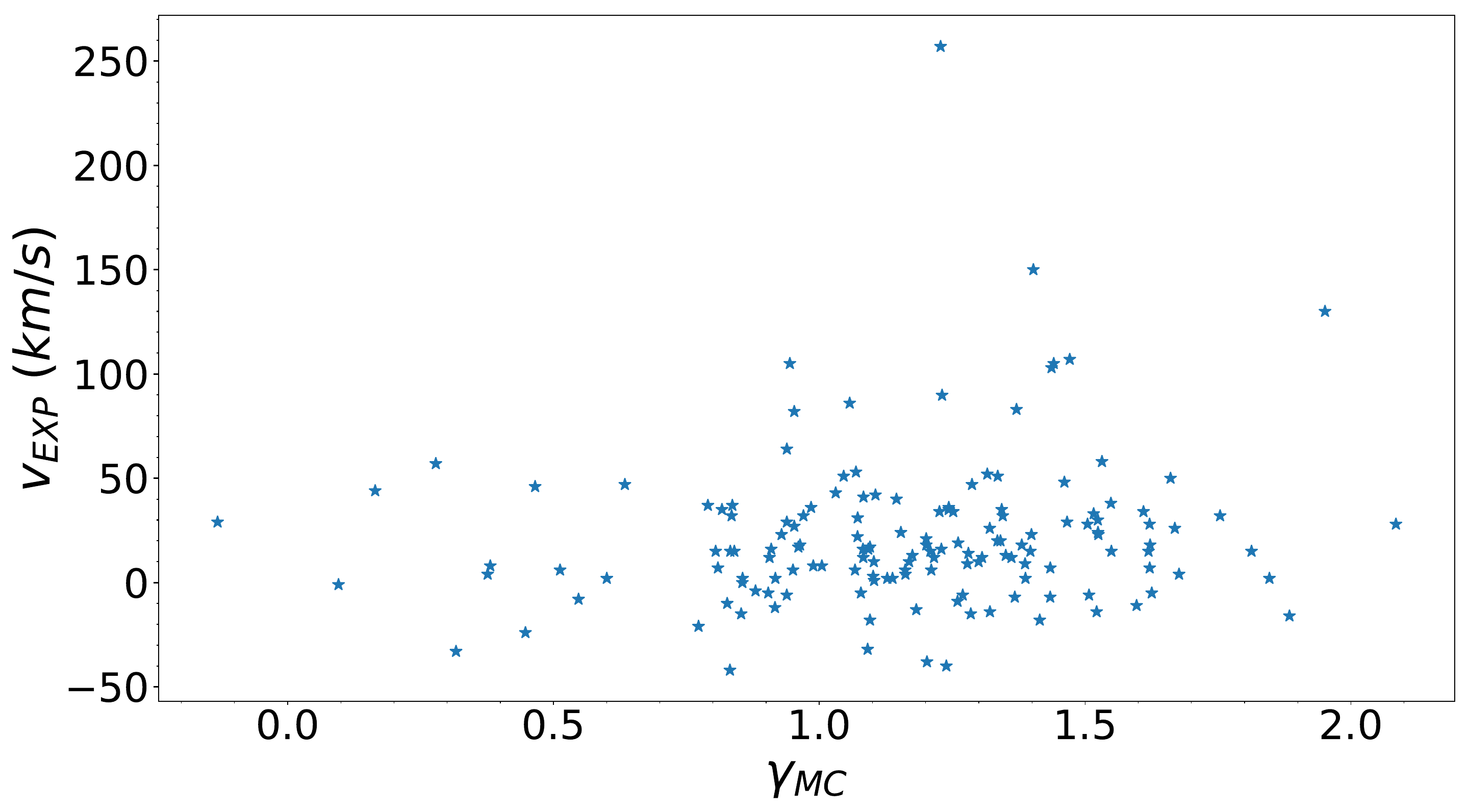}
	
	\caption{Figure showing the plot between the proton polytropic index inside MC ($\gamma_{\rm MC}$) and the cross-sectional expansion speed of MC ($v_{\rm EXP}$). This plot also shows that there is no correlation between these two parameters, suggesting that the cross-sectional expansion of near-Earth CMEs is unaffected by their bulk thermodynamic properties.} 
\label{Fig: vexp}
\end{figure}

However, we find that $r$ between $\gamma_{\rm MC}$ and $v_{\rm EXP}$ is only 0.14 with a p-value of 0.1, suggesting that there is no correlation between $\gamma_{\rm MC}$ and $v_{\rm EXP}$. This result is also evident from Fig~\ref{Fig: vexp}. The outcome therefore suggests that the cross-sectional expansion of MCs at 1 AU is not caused by the volumetric change by the polytropic process of MC plasma. Contextually, we also note that the excess pressure of the MC plasma with respect to the surrounding ambient solar wind does not contribute to $v_{\rm EXP}$ at 1 AU \citep{2023Debesh2}.
\begin{figure}[h!]

\includegraphics[width=0.90\textwidth , scale=0.9]{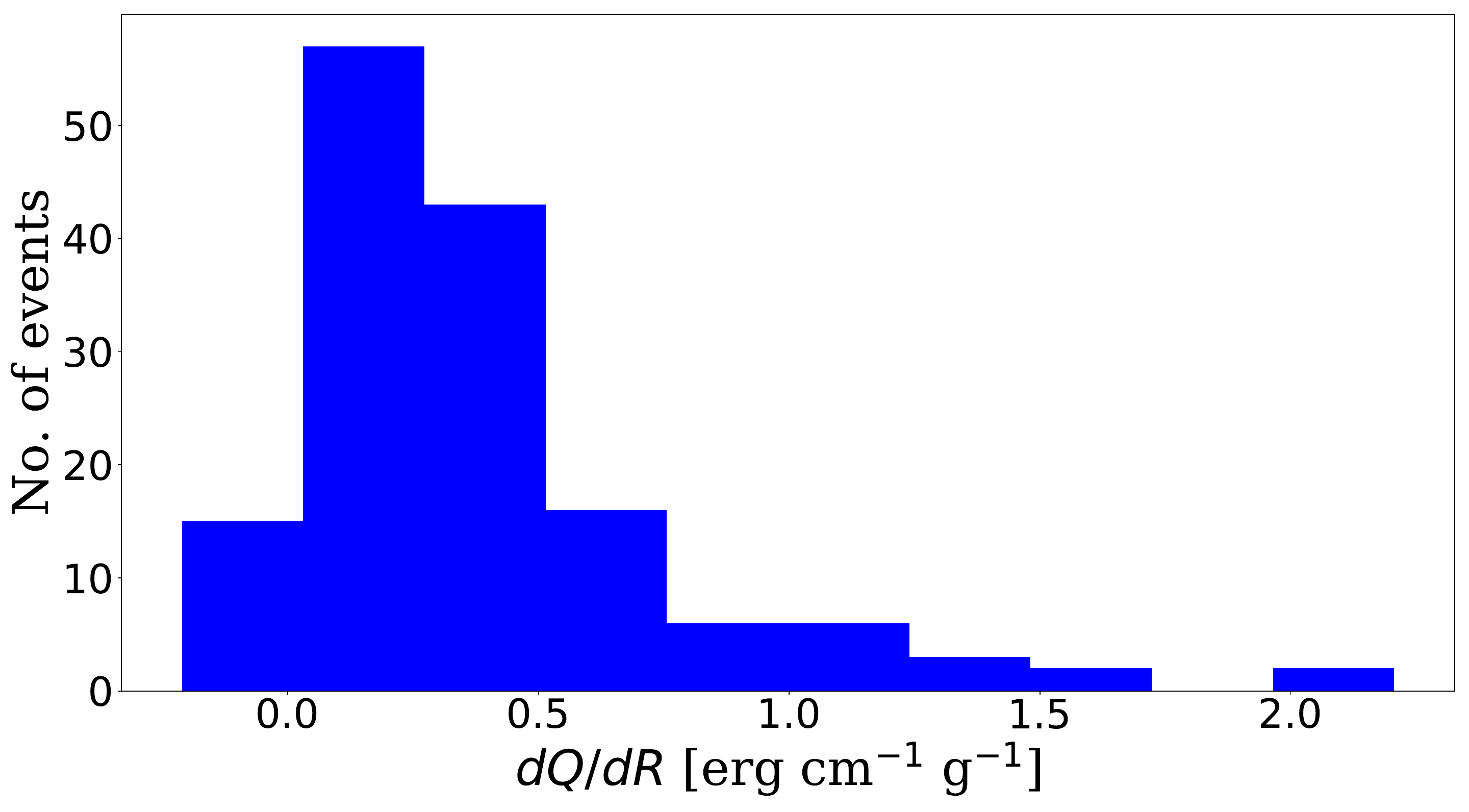}

\caption{Figure showing the histogram of the heating gradient per unit mass ($dQ/dR$, erg cm$^{-1}$ g$^{-1}$, Eq~\ref{eq dqdr}) of the protons of near-Earth MC plasma.} 
\label{Fig: dqdR}
\end{figure}
\subsection{Heating gradient}
\label{S - heating gradient}
From the findings of our study, it is clear that protons in both the MC and sheath plasma at the location of the Wind spacecraft follow a sub-adiabatic process ($\gamma < 5/3$).  A sub-adiabatic transition implies positive heating ($dQ > 0$). Therefore, the system needs additional heating. MCs are often assumed to be magnetically connected with the solar corona \citep{1993ChenGarren}. This additional heating of the MC plasma can be broadly catagorized into two types: i) The protons can be heated via a constant heat supply from the corona. ii) There is some additional local heating (such as turbulent heating) that enables the protons to maintain their temperature at 1 AU. In this section, we are aiming to address the first mechanism by quantifying the radial heating gradient of protons of MC plasma coming from the solar corona, at 1 AU. The radial heating gradient per unit mass ($dQ/dR$, erg cm$^{-1}$ g$^{-1}$) for protons in a non-adiabatic process can be written as (following Eq 15 of \cite{2023livadiotis})
\begin{equation}
\label{eq dqdr}
\frac{dQ}{dR} = \frac{a_{\rm n}d_{\rm f}}{2} \frac{k_{\rm B} T}{m_{\rm p}R} (\gamma_{a} - \gamma)
\end{equation}
where, $R$ is the radial distance from the Sun, $d_{\rm f}$ is the degrees of freedom per proton, $m_{\rm p}$ is the proton mass in g, $k_{\rm B}$ is the Boltzmann's constant, and $a_{\rm n}$ represents the power law index of the radial variation of proton density which is assumed as $n_{\rm p}(R) = n_{\rm p}(R_0) (R/R_0)^{-a_{\rm n}}$ \citep{2016Elliott,2017McComas}. We note that Eq~\ref{eq dqdr} is used for solar wind. However, for the time being, we are using it here for the analysis of MC plasma. In this study, we assume that $a_{\rm n}$ and $\gamma$ vary slowly with distance. Using the solar wind electron density variation model by \cite{1998leblanc1998} and assuming the electron density to be equal to the proton density \citep{2021SasiApJ}, we note that $n_{\rm p}(R) \propto R^{-2}$ from a few tens of $R_{\odot}$ to and beyond 1 AU \citep{1998leblanc1998}. Therefore, we use $a_{\rm n} = 2$. Assuming three degrees of freedom per proton, we use $d_{\rm f} = 3$. Taking into account all of these, we determine the heating gradient per unit mass at $R = 1$ AU for the MCs listed in Table~\ref{S - Table A}. Here, we use the temporal average of the temperature measured at 1 AU inside the MC and observed by the Wind spacecraft. The results are described in the histogram of Fig~\ref{Fig: dqdR}. The mean, median, and mpv of the histogram are 0.39, 0.30, and 0.21, respectively. We note that $dQ/dR \propto R^{-1}$ (Eq~\ref{eq dqdr}), which implies that the ICME gets less energy per distance as it propagates far from the Sun. This shows that by the time the ICME reaches close to the Earth, the heat that is supplied from the corona per unit mass per unit distance is $< 1$ erg. Considering the mpv, the radial heating gradient per unit mass of $\approx 0.21$ erg cm$^{-1}$ g$^{-1}$, can reach 1 AU to maintain the proton temperature in the MC plasma. Multiplying it with the typical ICME speed at 1 AU, $v_{\rm ICME} \approx 400$ km/s, we get that the amount of energy transferred from the corona to the MC at 1 AU is $\approx 820$ J kg$^{-1}$ s$^{-1}$. \cite{2006Liu} showed that the average required heat for ICMEs at 1 AU is 2550 J kg$^{-1}$ s$^{-1}$, of which 900 J kg$^{-1}$ s$^{-1}$ is accounted for by protons. This result is in general agreement with our estimate of $\approx 820$ J kg$^{-1}$ s$^{-1}$.

\section{Conclusion and Discussion}
\label{S - conclusion}
The problem of energy dissipation in ICME, sheath and solar wind plasma is often addressed by assuming that the plasma follows a polytropic equation of state, leading us to estimate the polytropic index, $\gamma$. The polytropic equation of state is essentially a simple and convenient substitute for the full energy equation. If $\gamma \approx 1$ (isothermal), it would suggest that the plasma parcel is thermally well-connected with the solar corona. On the other hand, for $\gamma \approx 5/3$ (adiabatic), it would imply that the plasma parcel needs an additional heating source to maintain its proton temperature close to the Earth. In this study, we use the Wind SWE and OMNI data to characterize the polytropic index of 150 well-observed MCs and 124 ICME sheaths associated with these events between 1995 and 2015 at 1 AU. Our main findings are the following: 

\begin{itemize}
	\item We find that $\gamma$ of the ambient solar wind (preceding the ICME event) at 1 AU is close to the adiabatic value ($\gamma \approx \gamma_a = 5/3$), while, inside the sheath plasma, $\gamma$ drops close to the isothermal value ($\gamma \approx \gamma_a = 1$). During the passage of the MC, $\gamma$ reaches an intermediate value (higher than the sheath, but lower than the ambient solar wind background). This high-low-high behavior of proton $\gamma$ is in agreement with the findings of \cite{2022dayeh}. We further note that the near-isothermal behavior of ICME sheaths does not depend on the sheath thickness at 1 AU. 
	
	\item  A polytrope is a macroscopic process that relates various key physical parameters like plasma density and temperature or pressure and volume. This leads us to investigate whether the change of these parameters via a polytropic transition affects turbulent density fluctuations in protons. Therefore, we estimate the correlation coefficients between polytropic indices inside and density modulation indices, and find that there is no correlation between them. This suggests that the macroscopic changes of MC and sheath plasma close to the Earth are independent of small-scale turbulent fluctuations of protons. This is in keeping with previous results which suggest that turbulent dissipation is not necessarily related to the polytropic processes in solar wind \citep{1995goldstein}. This implies that turbulent fluctuations are intrinsic properties of the plasma and independent of the polytropic changes.
	
	\item Additionally, we investigate if the polytropic changes affect the cross-sectional speed of MC near the Earth. However, we do not find any correlation between the cross-sectional expansion speed of MCs and $\gamma$ close to the Earth. It implies that this speed is independent of the volumetric changes associated with a polytropic process.

	\item We find that protons in MCs follow a sub-adiabatic behavior close to the Earth, suggesting positive plasma heating. The required heating of the MC plasma can be either via thermal conduction from the solar corona or by local turbulent heating. In this paper, we investigate how much heat per unit mass per unit distance travels radially from the solar corona to the protons of the MC plasma at 1 AU via thermal conduction. Our findings show that an amount of $\approx 0.21$ erg cm$^{-1}$ g$^{-1}$ energy, 
	can reach 1 AU to maintain the temperature of protons in MC plasma via thermal conduction. We note that our estimated radial heating ($\approx 820$ J kg$^{-1}$ s$^{-1}$) is in agreement with the required heating budget from \cite{2006Liu} (900 J kg$^{-1}$ s$^{-1}$). However, we also note that the transfer of heat via thermal conduction through electron-proton collisions is inadequate to satisfy the required heating budget for protons in the MC plasma at 1 AU \citep{saikat_thesis}. Therefore, the exact mechanism of thermal conduction in this context needs further investigation which is outside of the scope of this paper.
	
\end{itemize}

\appendix

\section{ICME list}

	\begin{table}[h!]
		\caption{
			The list of the 152 Wind ICME events enlisted in \cite{2023Debesh}. The arrival date and time of the ICME at the position of Wind measurement and the arrival and departure dates \& times of the associated magnetic clouds (MCs) are taken from Wind ICME catalogue (\url{https://wind.nasa.gov/ICMEindex.php}). Fr events indicate MCs with a single magnetic field rotation between $90^{\circ}$ and $180^{\circ}$ and F+ events indicate MCs with a single magnetic field rotation greater than $180^{\circ}$ \citep{2018NC}. The 14 events marked with and asterisk (*) coincide with the near earth counterparts of 14 CMEs listed in \citep{2017Nishtha}. }
		\label{S - Table A}

		\begin{center}
			
			\begin{tabular}{cccclccc}
				
				\hline\hline
				CME        & CME Arrival date & MC start  & MC end       & Flux rope \\
				event      & and time[UT]   &  date and   & date and     & type   \\
				number    &  (1AU)         &  time [UT]   &   time [UT]   &      \\
				\hline
				
				1    &  1995 03 04 , 00:36 & 1995 03 04 , 11:23 & 1995 03 05 , 03:06 & Fr \\
				2    &  1995 04 03 , 06:43 & 1995 04 03 , 12:45 & 1995 04 04 , 13:25 & F+ \\
				3    &  1995 06 30 , 09:21 & 1995 06 30 , 14:23 & 1995 07 02 , 16:47 & Fr \\
				4    &  1995 08 22 , 12:56 & 1995 08 22 , 22:19 & 1995 08 23 , 18:43 & Fr \\
				5    &  1995 09 26 , 15:57 & 1995 09 27 , 03:36 & 1995 09 27 , 21:21 & Fr \\
				6    &  1995 10 18 , 10:40 & 1995 10 18 , 19:11 & 1995 10 20 , 02:23 & Fr \\
				7    &  1996 02 15 , 15:07 & 1996 02 15 , 15:07 & 1996 02 16 , 08:59 & F+ \\
				8    &  1996 04 04 , 11:59 & 1996 04 04 , 11:59 & 1996 04 04 , 21:36 & Fr \\ 
				9    &  1996 05 16 , 22:47 & 1996 05 17 , 01:36 & 1996 05 17 , 11:58 & F+ \\
				10   &  1996 05 27 , 14:45 & 1996 05 27 , 14:45 & 1996 05 29 , 02:22 & Fr \\
				11   &  1996 07 01 , 13:05 & 1996 07 01 , 17:16 & 1996 07 02 , 10:17 & Fr \\
				12   &  1996 08 07 , 08:23 & 1996 08 07 , 11:59 & 1996 08 08 , 13:12 & Fr \\
				13   &  1996 12 24 , 01:26 & 1996 12 24 , 03:07 & 1996 12 25 , 11:44 & F+ \\ 
				14   &  1997 01 10 , 00:52 & 1997 01 10 , 04:47 & 1997 01 11 , 03:36 & F+ \\
				15   &  1997 04 10 , 17:02 & 1997 04 11 , 05:45 & 1997 04 11 , 19:10 & Fr \\
				16   &  1997 04 21 , 10:11 & 1997 04 21 , 11:59 & 1997 04 23 , 07:11 & F+ \\
				17   &  1997 05 15 , 01:15 & 1997 05 15 , 10:00 & 1997 05 16 , 02:37 & F+ \\
				18   &  1997 05 26 , 09:09 & 1997 05 26 , 15:35 & 1997 05 28 , 00:00 & Fr \\
				19   &  1997 06 08 , 15:43 & 1997 06 09 , 06:18 & 1997 06 09 , 23:01 & Fr \\
				20   &  1997 06 19 , 00:00 & 1997 06 19 , 05:31 & 1997 06 20 , 22:29 & Fr \\
				21   &  1997 07 15 , 03:10 & 1997 07 15 , 06:48 & 1997 07 16 , 11:16 & F+ \\
				22   &  1997 08 03 , 10:10 & 1997 08 03 , 13:55 & 1997 08 04 , 02:23 & Fr \\
				23   &  1997 08 17 , 01:56 & 1997 08 17 , 06:33 & 1997 08 17 , 20:09 & Fr \\
				24   &  1997 09 02 , 22:40 & 1997 09 03 , 08:38 & 1997 09 03 , 20:59 & Fr \\
				25   &  1997 09 18 , 00:30 & 1997 09 18 , 04:07 & 1997 09 19 , 23:59 & F+ \\
				26   &  1997 10 01 , 11:45 & 1997 10 01 , 17:08 & 1997 10 02 , 23:15 & Fr \\
				27   &  1997 10 10 , 03:08 & 1997 10 10 , 15:33 & 1997 10 11 , 22:00 & F+ \\
				28   &  1997 11 06 , 22:25 & 1997 11 07 , 06:00 & 1997 11 08 , 22:46 & F+ \\
				29   &  1997 11 22 , 09:12 & 1997 11 22 , 17:31 & 1997 11 23 , 18:43 & F+ \\
				30   &  1997 12 30 , 01:13 & 1997 12 30 , 09:35 & 1997 12 31 , 08:51 & Fr \\
				\hline
				
			\end{tabular}
		\end{center}
	\end{table}


	\begin{table}
		\begin{center}
			\begin{tabular}{cccclccc}
				\hline \hline
				CME        & CME Arrival date & MC start  & MC end       & Flux rope \\
				event      & and time[UT]   &  date and   & date and     & type   \\
				number    &  (1AU)         &  time [UT]   &   time [UT]   &      \\
				\hline
				
				31   &  1998 01 06 , 13:29 & 1998 01 07 , 02:23 & 1998 01 08 , 07:54 & F+ \\
				32   &  1998 01 28 , 16:04 & 1998 01 29 , 13:12 & 1998 01 31 , 00:00 & F+ \\
				33   &  1998 03 25 , 10:48 & 1998 03 25 , 14:23 & 1998 03 26 , 08:57 & Fr \\
				34   &  1998 03 31 , 07:11 & 1998 03 31 , 11:59 & 1998 04 01 , 16:18 & Fr \\
				35   &  1998 05 01 , 21:21 & 1998 05 02 , 11:31 & 1998 05 03 , 16:47 & Fr \\
				36   &  1998 06 02 , 10:28 & 1998 06 02 , 10:28 & 1998 06 02 , 09:16 & Fr \\
				37   &  1998 06 24 , 10:47 & 1998 06 24 , 13:26 & 1998 06 25 , 22:33 & F+ \\
				38   &  1998 07 10 , 22:36 & 1998 07 10 , 22:36 & 1998 07 12 , 21:34 & F+ \\
				39   &  1998 08 19 , 18:40 & 1998 08 20 , 08:38 & 1998 08 21 , 20:09 & F+ \\
				40   &  1998 10 18 , 19:30 & 1998 10 19 , 04:19 & 1998 10 20 , 07:11 & F+ \\
				
				41   &  1999 02 11 , 17:41 & 1999 02 11 , 17:41 & 1999 02 12 , 03:35 & Fr \\
				42   &  1999 07 02 , 00:27 & 1999 07 03 , 08:09 & 1999 07 05 , 13:13 & Fr \\
				43   &  1999 09 21 , 18:57 & 1999 09 21 , 18:57 & 1999 09 22 , 11:31 & Fr \\
				44   &  2000 02 11 , 23:34 & 2000 02 12 , 12:20 & 2000 02 13 , 00:35 & Fr \\
				45   &  2000 02 20 , 21:03 & 2000 02 21 , 14:24 & 2000 02 22 , 13:16 & Fr \\
				
				46   &  2000 03 01 , 01:58 & 2000 03 01 , 03:21 & 2000 03 02 , 03:07 & Fr \\
				47   &  2000 07 01 , 07:12 & 2000 07 01 , 07:12 & 2000 07 02 , 03:34 & Fr \\
				48   &  2000 07 11 , 22:35 & 2000 07 11 , 22:35 & 2000 07 13 , 04:33 & Fr \\
				49   &  2000 07 28 , 06:38 & 2000 07 28 , 14:24 & 2000 07 29 , 10:06 & F+ \\
				50   &  2000 09 02 , 23:16 & 2000 09 02 , 23:16 & 2000 09 03 , 22:32 & Fr \\
				51   &  2000 10 03 , 01:02 & 2000 10 03 , 09:36 & 2000 10 05 , 03:34 & F+ \\
				52   &  2000 10 12 , 22:33 & 2000 10 13 , 18:24 & 2000 10 14 , 19:12 & Fr \\
				53   &  2000 11 06 , 09:30 & 2000 11 06 , 23:05 & 2000 11 07 , 18:05 & Fr \\
				54   &  2000 11 26 , 11:43 & 2000 11 27 , 09:30 & 2000 11 28 , 09:36 & Fr \\
				55   &  2001 04 21 , 15:29 & 2001 04 22 , 00:28 & 2001 04 23 , 01:11 & Fr \\
				56   &  2001 10 21 , 16:39 & 2001 10 22 , 01:17 & 2001 10 23 , 00:47 & Fr \\
				57   &  2001 11 24 , 05:51 & 2001 11 24 , 15:47 & 2001 11 25 , 13:17 & Fr \\
				58   &  2001 12 29 , 05:16 & 2001 12 30 , 03:24 & 2001 12 30 , 19:10 & Fr \\
				59   &  2002 02 28 , 05:06 & 2002 02 28 , 19:11 & 2002 03 01 , 23:15 & Fr \\
				60   &  2002 03 18 , 13:14 & 2002 03 19 , 06:14 & 2002 03 20 , 15:36 & Fr \\
				61   &  2002 03 23 , 11:24 & 2002 03 24 , 13:11 & 2002 03 25 , 21:36 & Fr \\
				62   &  2002 04 17 , 11:01 & 2002 04 17 , 21:36 & 2002 04 19 , 08:22 & F+ \\
				63   &  2002 07 17 , 15:56 & 2002 07 18 , 13:26 & 2002 07 19 , 09:35 & Fr \\
				64   &  2002 08 18 , 18:40 & 2002 08 19 , 19:12 & 2002 08 21 , 13:25 & Fr \\
				65   &  2002 08 26 , 11:16 & 2002 08 26 , 14:23 & 2002 08 27 , 10:47 & Fr \\
				
				\hline
				
			\end{tabular}
		\end{center}
	\end{table}
	


	\begin{table}
		\begin{center}
			\begin{tabular}{cccclccc}
				\hline \hline
				CME        & CME Arrival date & MC start  & MC end       & Flux rope \\
				event      & and time[UT]   &  date and   & date and     & type   \\
				number    &  (1AU)         &  time [UT]   &   time [UT]   &      \\
				\hline

				66   &  2002 09 30 , 07:54 & 2002 09 30 , 22:04 & 2002 10 01 , 20:08 & F+ \\
				67   &  2002 12 21 , 03:21 & 2002 12 21 , 10:20 & 2002 12 22 , 15:36 & Fr \\
				68   &  2003 01 26 , 21:43 & 2003 01 27 , 01:40 & 2003 01 27 , 16:04 & Fr \\
				69   &  2003 02 01 , 13:06 & 2003 02 02 , 19:11 & 2003 02 03 , 09:35 & Fr \\
				70   &  2003 03 20 , 04:30 & 2003 03 20 , 11:54 & 2003 03 20 , 22:22 & Fr \\
				71   &  2003 06 16 , 22:33 & 2003 06 16 , 17:48 & 2003 06 18 , 08:18 & Fr \\
				72   &  2003 08 04 , 20:23 & 2003 08 05 , 01:10 & 2003 08 06 , 02:23 & Fr \\
				73   &  2003 11 20 , 08:35 & 2003 11 20 , 11:31 & 2003 11 21 , 01:40 & Fr \\
				74   &  2004 04 03 , 09:55 & 2004 04 04 , 01:11 & 2004 04 05 , 19:11 & F+ \\
				75   &  2004 09 17 , 20:52 & 2004 09 18 , 12:28 & 2004 09 19 , 16:58 & Fr \\
				
				76   &  2005 05 15 , 02:10 & 2005 05 15 , 05:31 & 2005 05 16 , 22:47 & F+ \\
				77   &  2005 05 20 , 04:47 & 2005 05 20 , 09:35 & 2005 05 22 , 02:23 & F+ \\
				78   &  2005 07 17 , 14:52 & 2005 07 17 , 14:52 & 2005 07 18 , 05:59 & Fr \\
				79   &  2005 10 31 , 02:23 & 2005 10 31 , 02:23 & 2005 10 31 , 18:42 & Fr \\
				80   &  2006 02 05 , 18:14 & 2006 02 05 , 20:23 & 2006 02 06 , 11:59 & F+ \\
				81   &  2006 09 30 , 02:52 & 2006 09 30 , 08:23 & 2006 09 30 , 22:03 & F+ \\
				82   &  2006 11 18 , 07:11 & 2006 11 18 , 07:11 & 2006 11 20 , 04:47 & Fr \\
				83   &  2007 05 21 , 22:40 & 2007 05 21 , 22:45 & 2007 05 22 , 13:25 & Fr \\
				84   &  2007 06 08 , 05:45 & 2007 06 08 , 05:45 & 2007 06 09 , 05:15 & Fr \\
				85   &  2007 11 19 , 17:22 & 2007 11 20 , 00:33 & 2007 11 20 , 11:31 & Fr \\

				86   &  2008 05 23 , 01:12 & 2008 05 23 , 01:12 & 2008 05 23 , 10:46 & F+ \\
				87   &  2008 09 03 , 16:33 & 2008 09 03 , 16:33 & 2008 09 04 , 03:49 & F+ \\
				88   &  2008 09 17 , 00:43 & 2008 09 17 , 03:57 & 2008 09 18 , 08:09 & Fr \\
				89   &  2008 12 04 , 11:59 & 2008 12 04 , 16:47 & 2008 12 05 , 10:47 & Fr \\
				90   &  2008 12 17 , 03:35 & 2008 12 17 , 03:35 & 2008 12 17 , 15:35 & Fr \\
				91   &  2009 02 03 , 19:21 & 2009 02 03 , 01:12 & 2009 02 04 , 19:40 & F+ \\
				92   &  2009 03 11 , 22:04 & 2009 03 12 , 01:12 & 2009 03 13 , 01:40 & F+ \\
				93   &  2009 04 22 , 11:16 & 2009 04 22 , 14:09 & 2009 04 22 , 20:37 & Fr \\
				94   &  2009 06 03 , 13:40 & 2009 06 03 , 20:52 & 2009 06 05 , 05:31 & Fr \\
				95   &  2009 06 27 , 11:02 & 2009 06 27 , 17:59 & 2009 06 28 , 20:24 & F+ \\ 
				
				96   &  2009 07 21 , 02:53 & 2009 07 21 , 04:48 & 2009 07 22 , 03:36 & Fr \\
				97   &  2009 09 10 , 10:19 & 2009 09 10 , 10:19 & 2009 09 10 , 19:26 & Fr \\
				
				98   &  2009 09 30 , 00:44 & 2009 09 30 , 06:59 & 2009 09 30 , 19:11 & Fr \\
				99   &  2009 10 29 , 01:26 & 2009 10 29 , 01:26 & 2009 10 29 , 23:45 & F+ \\
				100   &  2009 11 14 , 10:47 & 2009 11 14 , 10:47 & 2009 11 15 , 11:45 & Fr \\

				\hline
				
			\end{tabular}
		\end{center}
	\end{table}
	


	\begin{table}
		\begin{center}
			\begin{tabular}{cccclccc}
				\hline\hline
				CME        & CME Arrival date & MC start  & MC end       & Flux rope \\
				event      & and time[UT]   &  date and   & date and     & type   \\
				number    &  (1AU)         &  time [UT]   &   time [UT]   &      \\
				\hline
				
				101   &  2009 12 12 , 04:47 & 2009 12 12 , 19:26 & 2009 12 14 , 04:47 & Fr \\
				102   &  2010 01 01 , 22:04 & 2010 01 02 , 00:14 & 2010 01 03 , 09:06 & Fr \\
				103   &  2010 02 07 , 18:04 & 2010 02 07 , 19:11 & 2010 02 09 , 05:42 & Fr \\
				104*   &  2010 03 23 , 22:29 & 2010 03 23 , 22:23 & 2010 03 24 , 15:36 & Fr \\
				105*   &  2010 04 05 , 07:55 & 2010 04 05 , 11:59 & 2010 04 06 , 16:48 & Fr \\
				106*   &  2010 04 11 , 12:20 & 2010 04 11 , 21:36 & 2010 04 12 , 14:12 & Fr \\
				107   &  2010 05 28 , 01:55 & 2010 05 29 , 19:12 & 2010 05 29 , 17:58 & Fr \\
				108*   &  2010 06 21 , 03:35 & 2010 06 21 , 06:28 & 2010 06 22 , 12:43 & Fr \\
				109*   &  2010 09 15 , 02:24 & 2010 09 15 , 02:24 & 2010 09 16 , 11:58 & Fr \\
				110*   &  2010 10 31 , 02:09 & 2010 10 30 , 05:16 & 2010 11 01 , 20:38 & Fr \\
				111   &  2010 12 19 , 00:35 & 2010 12 19 , 22:33 & 2010 12 20 , 22:14 & F+ \\
				112   &  2011 01 24 , 06:43 & 2011 01 24 , 10:33 & 2011 01 25 , 22:04 & F+ \\
				113*   &  2011 03 29 , 15:12 & 2011 03 29 , 23:59 & 2011 04 01 , 14:52 & Fr \\
				114   &  2011 05 28 , 00:14 & 2011 05 28 , 05:31 & 2011 05 28 , 22:47 & F+ \\
				115   &  2011 06 04 , 20:06 & 2011 06 05 , 01:12 & 2011 06 05 , 18:13 & Fr \\
				116   &  2011 07 03 , 19:12 & 2011 07 03 , 19:12 & 2011 07 04 , 19:12 & Fr \\
				117*   &  2011 09 17 , 02:57 & 2011 09 17 , 15:35 & 2011 09 18 , 21:07 & Fr \\
				118   &  2012 02 14 , 07:11 & 2012 02 14 , 20:52 & 2012 02 16 , 04:47 & Fr \\
				119   &  2012 04 05 , 14:23 & 2012 04 05 , 19:41 & 2012 04 06 , 21:36 & Fr \\
				120   &  2012 05 03 , 00:59 & 2012 05 04 , 03:36 & 2012 05 05 , 11:22 & Fr \\
				121   &  2012 05 16 , 12:28 & 2012 05 16 , 16:04 & 2012 05 18 , 02:11 & Fr \\
				122   &  2012 06 11 , 02:52 & 2012 06 11 , 11:31 & 2012 06 12 , 05:16 & Fr \\
				123*   &  2012 06 16 , 09:03 & 2012 06 16 , 22:01 & 2012 06 17 , 11:23 & F+ \\
				124*   &  2012 07 14 , 17:39 & 2012 07 15 , 06:14 & 2012 07 17 , 03:22 & Fr \\
				125   &  2012 08 12 , 12:37 & 2012 08 12 , 19:12 & 2012 08 13 , 05:01 & Fr \\
				126   &  2012 08 18 , 03:25 & 2012 08 18 , 19:12 & 2012 08 19 , 08:22 & Fr \\
				127*   &  2012 10 08 , 04:12 & 2012 10 08 , 15:50 & 2012 10 09 , 17:17 & Fr \\
				128   &  2012 10 12 , 08:09 & 2012 10 12 , 18:09 & 2012 10 13 , 09:14 & Fr \\
				129*   &  2012 10 31 , 14:28 & 2012 10 31 , 23:35 & 2012 11 02 , 05:21 & F+ \\
				130*   &  2013 03 17 , 05:21 & 2013 03 17 , 14:09 & 2013 03 19 , 16:04 & Fr \\

				131*   &  2013 04 13 , 22:13 & 2013 04 14 , 17:02 & 2013 04 17 , 05:30 & F+ \\
				132   &  2013 04 30 , 08:52 & 2013 04 30 , 12:00 & 2013 05 01 , 07:12 & Fr \\
				133   &  2013 05 14 , 02:23 & 2013 05 14 , 06:00 & 2013 05 15 , 06:28 & Fr \\
				134   &  2013 06 06 , 02:09 & 2013 06 06 , 14:23 & 2013 06 08 , 00:00 & F+ \\
				135   &  2013 06 27 , 13:51 & 2013 06 28 , 02:23 & 2013 06 29 , 11:59 & Fr \\
				\hline
				
			\end{tabular}
		\end{center}
	\end{table}   


	\begin{table}
		\begin{center}
			\begin{tabular}{cccclccc}
				\hline\hline
				CME        & CME Arrival date & MC start  & MC end       & Flux rope \\
				event      & and time[UT]   &  date and   & date and     & type   \\
				number    &  (1AU)         &  time [UT]   &   time [UT]   &      \\
				\hline
				
				136   &  2013 09 01 , 06:14 & 2013 09 01 , 13:55 & 2013 09 02 , 01:56 & Fr \\
				137   &  2013 10 30 , 18:14 & 2013 10 30 , 18:14 & 2013 10 31 , 05:30 & Fr \\
				138   &  2013 11 08 , 21:07 & 2013 11 08 , 23:59 & 2013 11 09 , 06:14 & Fr \\
				139   &  2013 11 23 , 00:14 & 2013 11 23 , 04:47 & 2013 11 23 , 15:35 & Fr \\
				140   &  2013 12 14 , 16:47 & 2013 12 15 , 16:47 & 2013 12 16 , 05:30 & Fr \\
				141   &  2013 12 24 , 20:36 & 2013 12 25 , 04:47 & 2013 12 25 , 17:59 & F+ \\
				142   &  2014 04 05 , 09:58 & 2014 04 05 , 22:18 & 2014 04 07 , 14:24 & Fr \\
				143   &  2014 04 11 , 06:57 & 2014 04 11 , 06:57 & 2014 04 12 , 20:52 & F+ \\
				144   &  2014 04 20, 10:20 & 2014 04 21 , 07:41 & 2014 04 22 , 06:12 & Fr \\
				145   &  2014 04 29 , 19:11 & 2014 04 29 , 19:11 & 2014 04 30 , 16:33 & Fr \\
				146   &  2014 06 29 , 04:47 & 2014 06 29 , 20:53 & 2014 06 30 , 11:15 & Fr \\
				147   &  2014 08 19 , 05:49 & 2014 08 19 , 17:59 & 2014 08 21 , 19:09 & F+ \\
				148   &  2014 08 26 , 02:40 & 2014 08 27 , 03:07 & 2014 08 27 , 21:49 & Fr \\
				149   &  2015 01 07 , 05:38 & 2015 01 07 , 06:28 & 2015 01 07 , 21:07 & F+ \\
				150   &  2015 09 07 , 13:05 & 2015 09 07 , 23:31 & 2015 09 09 , 14:52 & F+ \\
				151   &  2015 10 06 , 21:35 & 2015 10 06 , 21:35 & 2015 10 07 , 10:03 & Fr \\
				152   &  2015 12 19 , 15:35 & 2015 12 20 , 13:40 & 2015 12 21 , 23:02 & Fr \\
				\hline
				
			\end{tabular}
		\end{center}
	\end{table}   

\section{The list of polytropic indices ($\gamma$)} 

\begin{table}[h!]
	\caption{ Polytopic index ($\gamma$) estimates from the fits between proton number density ($n_{\rm p}$) and temperature ($T$) inside MCs, sheaths, and solar wind backgrounds (BG1 and BG2). For simplicity, we keep the event numbers according to the Table~\ref{S - Table A}. We note that we have studied 150 MCs, and backgrounds (BG1 and BG2), and 124 sheaths in this paper.}
	\label{S - Table B}

	\begin{center}
		
		\begin{tabular}{cccccccc}
			
			\hline\hline
			CME        & $\gamma$ & $\gamma$  &  $\gamma$ & $\gamma$ \\
			event      & inside MC   &  inside sheath  & inside BG1     & inside BG2   \\
			number    &   &    &    &      \\
			\hline
			
			1    &  $1.10 \pm 0.11$ & $1.25 \pm 0.61$ & $1.71 \pm 0.07$ & $1.38 \pm 0.90$ \\
			2    &  $0.88 \pm 0.06$ & $0.90 \pm 0.14$ & $1.84 \pm 0.21$ & $1.78 \pm 0.26$ \\
			3    &  $0.96 \pm 0.10$ & $0.95 \pm 0.71$ & $1.55 \pm 0.41$ & $1.42 \pm 0.50$ \\
			4    &  $1.23 \pm 0.32$ & $0.80 \pm 0.25$ & $1.69 \pm 0.42$ & $1.52 \pm 0.17$ \\
			5    & $0.32 \pm 0.18$ & $0.73 \pm 0.20$ & $1.92 \pm 0.52$ & $1.40 \pm 0.23$ \\
			6    &  $0.39 \pm 0.11$ & $1.15 \pm 0.76$ & $1.22 \pm 0.30$ & $1.34 \pm 0.32$ \\
			7    &  $0.95 \pm 0.21$ & & $0.92 \pm 0.36$ & $1.63 \pm 0.53$ \\
			8    &  $0.52 \pm 0.24$ & & $1.72 \pm 0.40$ & $1.95 \pm 0.70$ \\ 
			9    &  $0.99 \pm 0.12$ & $1.49 \pm 0.07$ & $1.34 \pm 0.13$ & $1.39 \pm 0.22$ \\
			10   &  $2.08 \pm 0.38$ & & $0.79 \pm 0.43$ & $1.61 \pm 0.24$ \\
			11   &  $1.09 \pm 0.19$ & $1.34 \pm 0.09$ & $1.22 \pm 0.23$ & $1.87 \pm 0.21$ \\
			12   &  $0.93 \pm 0.08$ & $1.11 \pm 0.07$ & $1.52 \pm 0.20$ & $1.69 \pm 0.21$ \\
			13   &  $0.79 \pm 0.09$ & $0.68 \pm 0.22$ & $1.59 \pm 0.26$ & $1.74 \pm 0.33$ \\ 
			14   &  $1.23 \pm 1.1$ & $0.93 \pm 0.10$ & $1.47 \pm 0.21$ & $1.56 \pm 0.30$ \\
			15   &  $1.11 \pm 0.08$ & $1.05 \pm 0.07$ & $1.51 \pm 0.19$ & $1.40 \pm 0.09$ \\
			16   &  $1.09 \pm 0.10$ & $0.64 \pm 0.47$ & $1.02 \pm 0.03$ & $1.05 \pm 0.08$ \\
			17   &  $1.21 \pm 0.02$ & $0.69 \pm 0.11$ & $1.29 \pm 0.07$ & $1.42 \pm 0.18$ \\
			18   &  $1.10 \pm 0.09$ & $0.88 \pm 0.10$ & $1.20 \pm 0.10$ & $1.31 \pm 0.14$ \\
			19   &  $1.01 \pm 0.22$ & $1.01 \pm 0.03$ & $1.77 \pm 0.40$ & $1.33 \pm 0.11$ \\
			20   &  $1.24 \pm 0.09$ & $0.57 \pm 0.15$ & $1.10 \pm 0.12$ & $1.10 \pm 0.12$ \\
			21   &  $0.96 \pm 0.13$ & $1.31 \pm 0.17$ & $1.49 \pm 0.21$ & $1.48 \pm 0.30$ \\
			22   &  $1.11 \pm 0.17$ & $0.52 \pm 0.27$ & $1.56 \pm 0.11$ & $1.52 \pm 0.25$ \\
			23   &  $1.35 \pm 0.10$ & $1.12 \pm 0.03$ & $1.34 \pm 0.16$ & $1.38 \pm 0.09$ \\
			24   &  $0.91 \pm 0.32$ & $0.79 \pm 0.30$ & $1.69 \pm 0.04$ & $1.68 \pm 0.11$ \\
			25   &  $1.25 \pm 0.08$ & $1.28 \pm 0.27$ & $0.79 \pm 0.03$ & $1.92 \pm 0.09$ \\
			26   &  $1.52 \pm 0.11$ & $1.07 \pm 0.12$ & $1.36 \pm 0.50$ & $1.34 \pm 0.61$ \\
			27   &  $1.34 \pm 0.30$ & $0.87 \pm 0.34$ & $1.47 \pm 0.16$ & $1.63 \pm 0.45$ \\
			28   &  $1.29 \pm 0.12$ & $0.89 \pm 0.28$ & $1.31 \pm 0.20$ & $1.57 \pm 0.71$ \\
			29   &  $1.55 \pm 0.09$ & $0.94 \pm 0.11$ & $1.74 \pm 0.17$ & $1.50 \pm 0.12$ \\
			30   &  $0.89 \pm 0.10$ & $0.97 \pm 0.15$ & $1.59 \pm 0.07$ & $1.60 \pm 0.09$ \\
			\hline
			
		\end{tabular}
	\end{center}
\end{table}


\begin{table}
	\begin{center}
		\begin{tabular}{cccccccc}
			\hline \hline
			
			CME        & $\gamma$ & $\gamma$  &  $\gamma$ & $\gamma$ \\
			event      & inside MC   &  inside sheath  & inside BG1     & inside BG2   \\
			number    &   &    &    &      \\
			\hline
			
			31   &  $1.75 \pm 0.06$ & $1.08 \pm 0.05$ & $1.66 \pm 0.09$ & $1.69 \pm 0.11$ \\
			32   &  $1.62 \pm 0.05$ & $1.10 \pm 0.20$ & $1.61 \pm 0.08$ & $1.65 \pm 0.14$ \\
			33   &  $1.16 \pm 0.08$ & $1.34 \pm 0.19$ & $1.54 \pm 0.20$ & $1.58 \pm 0.11$ \\
			34   &  $1.22 \pm 0.03$ & $1.09 \pm 0.09$ & $1.59 \pm 0.11$ & $1.71 \pm 0.25$ \\
			35   & $1.44 \pm 0.24$  & $1.13 \pm 0.32$ & $1.64 \pm 0.50$ & $1.61 \pm 0.35$ \\
			36   &  $0.92 \pm 0.11$ & & $1.23 \pm 0.12$ & $1.28 \pm 0.21$ \\
			37   &  $1.66 \pm 0.05$ & $0.99 \pm 0.04$ & $1.22 \pm 0.04$ & $1.45 \pm 0.34$ \\
			38   &  $0.85 \pm 0.10$ &  & $1.22 \pm 0.11$ & $1.30 \pm 0.09$ \\
			39   &  $0.96 \pm 0.08$ & $1.50 \pm 0.38$ & $1.59 \pm 0.13$ & $1.91 \pm 0.45$ \\
			40   &  $1.62 \pm 0.07$ & $1.24 \pm 0.25$ & $1.63 \pm 0.11$ & $1.65 \pm 0.42$ \\
			
			41   &  $1.43 \pm 0.31$ & & $1.38 \pm 0.22$ & $1.45 \pm 0.09$ \\
			42   &  $1.09 \pm 0.10$ & $0.86 \pm 0.11$ & $0.90 \pm 0.11$ & $1.01 \pm 0.21$ \\
			43   &  $1.11 \pm 0.18$ & & $1.72 \pm 0.12$ & $1.65 \pm 0.34$ \\
			44   &  $1.18 \pm 0.07$ & $0.97 \pm 0.09$& $1.62 \pm 0.03$ & $1.59 \pm 0.04$ \\
			45   & $1.34 \pm 0.18$  & $0.95 \pm 0.13$& $1.48 \pm 0.31$ & $1.49 \pm 0.15$ \\
			
			46   & $1.05 \pm 0.04$ & $1.52 \pm 0.25$ & $1.63 \pm 0.52$ & $1.62 \pm 0.38$ \\
			47   &  $1.60 \pm 0.10$ & & $1.53 \pm 0.03$ & $1.38 \pm 0.08$ \\
			48   & $0.84 \pm 0.04$ & & $1.30 \pm 0.02$ & $1.26 \pm 0.82$ \\
			49   &  $1.62 \pm 0.43$ & $1.16 \pm 0.19$ & $1.70 \pm 0.13$ & $1.40 \pm 0.15$ \\
			50   & $1.61 \pm 0.03$ & & $1.53 \pm 0.40$ & $1.31 \pm 0.11$ \\
			51   & $1.62 \pm 0.09$ & $1.19 \pm 0.11$ & $1.50 \pm 0.10$ & $1.19 \pm 0.38$ \\
			52   & $0.96 \pm 0.12$ & $1.02 \pm 0.08$ & $1.17 \pm 0.10$ & $1.20 \pm 0.51$ \\
			53   &  $1.53 \pm 0.15$ & $0.80 \pm 0.10$ & $1.08 \pm 0.11$ & $1.39 \pm 0.20$ \\
			54   &  $1.46 \pm 0.18$ & $1.51 \pm 0.21$ & $1.49 \pm 0.22$ & $1.60 \pm 0.09$ \\
			55   &  $1.47 \pm 0.19$ & $1.05 \pm 0.11$ & $1.09 \pm 0.38$ & $1.83 \pm 0.62$ \\
			56   & $1.23 \pm 0.08$ & $0.96 \pm 0.14$ & $1.91 \pm 0.18$ & $1.67 \pm 0.32$ \\
			57   &  $1.40 \pm 0.06$ & $1.40 \pm 0.18$ & $1.45 \pm 0.17$ & $1.69 \pm 0.26$ \\
			58   &  $0.48 \pm 0.29$ & $1.36 \pm 0.20$ & $1.52 \pm 0.16$ & $1.42 \pm 0.10$ \\
			59   &  $0.81 \pm 0.16$ & $1.28 \pm 0.13$ & $1.58 \pm 0.12$ & $1.48 \pm 0.20$ \\
			60   & $1.32 \pm 0.09$ & $1.17 \pm 0.09$ & $1.51 \pm 0.11$ & $1.68 \pm 0.15$ \\
			61   & $1.39 \pm 0.12$ & $0.98 \pm 0.11$ & $1.48 \pm 0.29$ & $1.54 \pm 0.13$ \\
			62   & $1.33 \pm 0.10$ & $1.65 \pm 0.45$ & $1.37 \pm 0.31$ & $1.69 \pm 0.08$ \\
			63   & &  & & \\
			64   & $1.03 \pm 0.07$ & $0.94 \pm 0.18$ & $1.42 \pm 0.60$ & $1.45 \pm 0.19$ \\
			65   &  $1.39 \pm 0.20$ & $1.05 \pm 0.06$& $1.62 \pm 0.11$  & $1.49 \pm 0.22$ \\
			
			\hline
			
		\end{tabular}
	\end{center}
\end{table}



	\begin{table}
		\begin{center}
			\begin{tabular}{cccccccc}
				\hline \hline
				CME        & $\gamma$ & $\gamma$  &  $\gamma$ & $\gamma$ \\
				event      & inside MC   &  inside sheath  & inside BG1     & inside BG2   \\
				number    &   &    &    &      \\
				\hline

				66   &  $1.52 \pm 0.18$ & $1.76 \pm 0.30$& $2.09 \pm 0.36$ & $1.42 \pm 0.31$ \\
				67   & $1.68 \pm 0.08$ & $1.31 \pm 0.31$& $1.76 \pm 0.22$ & $1.72 \pm 0.24$ \\
				68   & $0.49 \pm 0.07$ & $1.16 \pm 0.19$& $1.53 \pm 0.48$ &  $1.66 \pm 0.71$ \\
				69   &  $0.88 \pm 0.07$ & $0.95 \pm 0.13$& $1.36 \pm 0.87$ & $1.66 \pm 0.68$ \\
				70   & $1.10 \pm 0.09$ & $1.69 \pm 0.39$& $1.59 \pm 0.22$ & $1.54 \pm 0.31$ \\
				71   & $1.11 \pm 0.11$ & $1.13 \pm 0.08$ & $1.54 \pm 0.71$ & $0.98 \pm 0.13$ \\
				72   & $1.17 \pm 0.08$ & $1.05 \pm 0.05$ & $1.58 \pm 0.33$ & $1.98 \pm 0.35$ \\
				73   &  &  &  &  \\
				74   & $0.96 \pm 0.10$ &  $1.03 \pm 0.09$ & $1.42 \pm 0.20$ & $1.55 \pm 0.18$ \\
				75   & $1.23 \pm 0.06$ & $0.57 \pm 0.32$ & $1.77 \pm 0.08$ & $1.70 \pm 0.28$ \\
				
				76   & $1.52 \pm 0.05$ & $1.05 \pm 0.10$ & $1.46 \pm 0.12$ & $1.50 \pm 0.69$ \\
				77   & $1.31 \pm 0.10$ & $1.09 \pm 0.21$ & $1.45 \pm 0.06$ & $1.48 \pm 0.33$ \\
				78   & $1.40 \pm 0.11$ & & $1.61 \pm 0.06$ & $1.48 \pm 0.23$ \\
				79   & $1.26 \pm 0.19$ & & $0.94 \pm 0.12$ & $0.83 \pm 0.48$ \\
				80   &  $1.20 \pm 0.13$ & $1.20 \pm 0.25$ & $1.60 \pm 0.21$ & $1.71 \pm 0.29$ \\
				81   & $1.32 \pm 0.13$ & $1.41 \pm 0.29$ & $1.41 \pm 0.11$ & $1.62 \pm 0.09$ \\
				82   & $1.51 \pm 0.04$ & & $1.67 \pm 0.17$ & $1.57 \pm 0.19$ \\
				83   & $1.62 \pm 0.11$ & $1.67 \pm 0.50$ & $1.50 \pm 0.08$ & $1.71 \pm 0.22$ \\
				84   &  $1.27 \pm 0.12$ & & $1.54 \pm 0.11$ & $1.98 \pm 0.20$ \\
				85   &  $1.85 \pm 0.57$ & $1.48 \pm 0.18$ & $1.72 \pm 0.10$ & $1.60 \pm 0.10$ \\

				86   & $1.20 \pm 0.11$ & & $0.94 \pm 0.06$ & $1.60 \pm 0.08$ \\
				87   & $0.79 \pm 0.29$ & & $1.53 \pm 0.19$ & $1.63 \pm 0.30$ \\
				88   & $0.29 \pm 0.09$ & $1.23 \pm 0.28$ & $1.45 \pm 0.22$ & $1.46 \pm 0.22$ \\
				89   & $1.21 \pm 0.03$ & $0.99 \pm 0.15$ & $1.30 \pm 0.12$ & $1.60 \pm 0.80$ \\
				90   & $1.36 \pm 0.10$ & & $1.30 \pm 0.15$ & $1.67 \pm 0.05$ \\
				91   & $1.09 \pm 0.10$ & $1.28 \pm 0.27$ & $1.55 \pm 0.16$ & $1.64 \pm 0.51$ \\
				92   & $1.42 \pm 0.18$ & $1.65 \pm 0.45$& $1.62 \pm 0.20$ & $1.63 \pm 0.31$ \\
				93   & $0.84 \pm 0.11$ & $1.43 \pm 0.45$& $1.59 \pm 0.06$ & $1.49 \pm 0.11$ \\
				94   & $1.13 \pm 0.08$ & $1.28 \pm 0.30$& $1.54 \pm 0.30$ & $1.52 \pm 0.28$ \\
				95   & $1.52 \pm 0.20$ & $1.01 \pm 0.04$ & $1.67 \pm 0.09$ & $1.72 \pm 0.13$ \\ 
				
				96   & $1.81 \pm 0.11$ & $1.62 \pm 0.38$ & $1.87 \pm 0.13$ & $0.94 \pm 0.61$ \\
				97   & $0.66 \pm 0.10$ &  & $1.42 \pm 0.22$ & $1.68 \pm 0.05$ \\
				
				98   & $1.21 \pm 0.04$ & $1.45 \pm 0.34$ & $1.36 \pm 0.24$ & $1.62 \pm 0.15$ \\
				99   & $0.93 \pm 0.20$ & & $1.63 \pm 0.16$ & $0.99 \pm 0.14$ \\
				100   & $1.28 \pm 0.08$ & & $1.79 \pm 0.26$ & $1.68 \pm 0.17$ \\

				\hline
				
			\end{tabular}
		\end{center}
	\end{table}
	


	\begin{table}
		\begin{center}
			\begin{tabular}{cccccccc}
				\hline \hline
				CME        & $\gamma$ & $\gamma$  &  $\gamma$ & $\gamma$ \\
				event      & inside MC   &  inside sheath  & inside BG1     & inside BG2   \\
				number    &   &    &    &      \\
				\hline
				
				101   & $0.87 \pm 0.10$ & $0.98 \pm 0.10$ & $1.34 \pm 0.03$ & $1.51 \pm 0.02$ \\
				102   & $1.31 \pm 0.03$ & $1.29 \pm 0.19$ & $1.58 \pm 0.11$ & $1.74 \pm 0.13$ \\
				103   & $1.24 \pm 0.05$ & $1.07 \pm 0.08$ & $1.53 \pm 0.10$ & $1.29 \pm 0.23$ \\
				104*   & $1.88 \pm 0.21$ & $1.40 \pm 0.32$ & $1.25 \pm 0.05$ & $1.39 \pm 0.10$ \\
				105*   & $1.95 \pm 0.15$ & $0.98 \pm 0.12$ & $1.43 \pm 0.08$ & $1.58 \pm 0.09$ \\
				106*   & $1.20 \pm 0.03$ & $1.09 \pm 0.10$ & $1.18 \pm 0.16$ & $1.65 \pm 0.04$ \\
				107   & $0.98 \pm 0.11$ & $1.01 \pm 0.08$ & $1.48 \pm 0.09$ & $1.70 \pm 0.02$ \\
				108*   & $1.67 \pm 0.10$ & $1.13 \pm 0.14$ & $1.54 \pm 0.20$ & $1.60 \pm 0.08$ \\
				109*   & $1.18 \pm 0.04$ & & $1.92 \pm 0.05$ & $1.44 \pm 0.35$ \\
				110*   & $1.09 \pm 0.10$ & $1.68 \pm 0.46$ & $1.87 \pm 0.05$ & $1.66 \pm 0.12$ \\
				111   & $0.98 \pm 0.16$ & $0.97 \pm 0.17$ & $1.62 \pm 0.11$ & $1.61 \pm 0.06$ \\
				112   & $0.99 \pm 0.09$ & $1.46 \pm 0.24$ & $1.44 \pm 0.10$ & $1.62 \pm 0.07$ \\
				113*   & $1.30 \pm 0.10$ & $0.80 \pm 0.21$ & $1.51 \pm 0.02$ & $1.50 \pm 0.09$ \\
				114   & $1.09 \pm 0.19$ & $1.11 \pm 0.13$ & $1.38 \pm 0.12$ & $1.87 \pm 0.54$ \\
				115   & $1.50 \pm 0.76$ & $1.80 \pm 0.30$ & $1.09 \pm 0.14$ & $1.58 \pm 0.28$ \\
				116   & $1.09 \pm 0.06$ & & $1.74 \pm 0.40$ & $1.59 \pm 0.12$ \\
				117*   & $0.87 \pm 0.08$ & $1.06 \pm 0.09$ & $1.73 \pm 0.41$ & $1.48 \pm 0.13$ \\
				118   & $0.97 \pm 0.04$ & $0.99 \pm 0.11$ & $1.59 \pm 0.09$ & $1.87 \pm 0.10$ \\
				119   & $1.34 \pm 0.04$ & $1.24 \pm 0.27$ & $1.84 \pm 0.11$ & $1.62 \pm 0.09$ \\
				120   & $0.87 \pm 0.05$ & $1.44 \pm 0.28$ & $1.69 \pm 0.02$ & $1.65 \pm 0.05$ \\
				121   &  $1.37 \pm 0.06$ & $0.98 \pm 0.15$ & $1.21 \pm 0.05$ & $1.63 \pm 0.08$ \\
				122   & $0.62 \pm 0.03$ & $0.97 \pm 0.19$ & $1.37 \pm 0.29$ & $1.70 \pm 0.31$ \\
				123*   & $1.27 \pm 0.82$ & $1.42 \pm 0.42$ & $1.22 \pm 0.03$ & $1.61 \pm 0.05$ \\
				124*   &  $0.99 \pm 0.09$ & $0.67 \pm 0.20$ & $1.34 \pm 0.06$ & $1.70 \pm 0.20$ \\
				125   & $1.14 \pm 0.08$ & $1.36 \pm 0.18$ & $1.94 \pm 0.13$ & $1.94 \pm 0.21$ \\
				126   & $1.17 \pm 0.04$ & $0.81 \pm 0.20$ & $1.23 \pm 0.05$ & $1.69 \pm 0.04$ \\
				127*   & $1.26 \pm 0.06$ & $0.91 \pm 0.10$ & $1.46 \pm 0.22$ & $1.35 \pm 0.10$ \\
				128   & $0.65 \pm 0.23$ & $1.05 \pm 0.04$ & $1.72 \pm 0.11$ & $1.78 \pm 0.11$ \\
				129*   & $0.90 \pm 0.15$ & $0.98 \pm 0.13$ & $1.87 \pm 0.07$ & $1.62 \pm 0.06$ \\
				130*   & $1.46 \pm 0.03$ & $0.93 \pm 0.17$ & $1.45 \pm 0.18$ & $1.45 \pm 0.10$ \\

				131*   & $1.33 \pm 0.10$ & $0.95 \pm 0.11$ & $1.31\pm 0.12$ & $1.51 \pm 0.14$ \\
				132   & $1.28 \pm 0.08$ & $0.89 \pm 0.13$ & $1.94 \pm 0.20$ & $1.52 \pm 0.17$ \\
				133   & $1.01 \pm 0.27$ & $0.87 \pm 0.30$ & $1.41 \pm 0.11$ & $1.37 \pm 0.31$ \\
				134   & $0.69 \pm 0.13$ & $0.99 \pm 0.09$ & $1.41 \pm 0.04$ & $1.39 \pm 0.15$ \\
				135   & $1.43 \pm 0.09$ & $1.30 \pm 0.57$ & $1.78 \pm 0.10$ & $1.80 \pm 0.21$ \\
				\hline
				
			\end{tabular}
		\end{center}
	\end{table}   


	\begin{table}
		\begin{center}
			\begin{tabular}{cccccccc}
				\hline \hline
				CME        & $\gamma$ & $\gamma$  &  $\gamma$ & $\gamma$ \\
				event      & inside MC   &  inside sheath  & inside BG1     & inside BG2   \\
				number    &   &    &    &      \\
				\hline
				
				136   & $1.08 \pm 0.02$ & $0.99 \pm 0.15$ & $1.73 \pm 0.10$ & $1.68 \pm 0.09$ \\
				137   &  $1.28 \pm 0.07$ & & $1.16 \pm 0.15$ & $1.51 \pm 0.10$ \\
				138   &  $1.13 \pm 0.13$ & $0.98 \pm 0.21$ & $1.82 \pm 0.11$ & $1.48 \pm 0.11$ \\
				139   & $0.83 \pm 0.08$ & $1.22 \pm 0.41$ & $1.43 \pm 0.12$ & $1.99 \pm 0.19$ \\
				140   & $1.16 \pm 0.10$ & $1.61 \pm 0.19$ & $1.97 \pm 0.16$ & $1.50 \pm 0.50$ \\
				141   & $1.52 \pm 0.27$ & $1.30 \pm 0.36$ & $1.72 \pm 0.10$ & $1.72 \pm 0.11$ \\
				142   &  $1.20 \pm 0.10$ & $1.05 \pm 0.09$ & $1.55 \pm 0.40$ & $1.61 \pm 0.06$ \\
				143   &  $1.16 \pm 0.08$ & & $1.70 \pm 0.06$ & $1.95 \pm 0.17$ \\
				144   & $1.15 \pm 0.09$ & $1.34 \pm 0.33$ & $1.31 \pm 0.49$ & $1.69 \pm 0.20$ \\
				145   & $1.08 \pm 0.12$ & & $1.21 \pm 0.13$ & $1.77 \pm 0.03$ \\
				146   &  $1.10 \pm 0.35$ & $1.56 \pm 0.47$ & $1.23 \pm 0.15$ & $1.44 \pm 0.11$ \\
				147   &  $1.55 \pm 0.12$ & $1.32 \pm 0.27$ & $1.63 \pm 0.05$ & $1.85 \pm 0.09$ \\
				148   & $0.87 \pm 0.16$ & $1.16 \pm 0.19$ & $1.54 \pm 0.12$ & $1.80 \pm 0.23$ \\
				149   & $1.38 \pm 0.21$ &$0.99 \pm 0.11$ & $1.41 \pm 0.14$ & $1.61 \pm 0.27$ \\
				150   & $1.37 \pm 0.09$ & $1.56 \pm 0.62$ & $1.55 \pm 0.12$ & $1.42 \pm 0.10$ \\
				151   & $1.24 \pm 0.06$ & & $1.50 \pm 0.25$ & $1.56 \pm 0.09$ \\
				152   & $1.40 \pm 0.10$ & $0.98 \pm 0.17$ & $1.54 \pm 0.21$ & $1.65 \pm 0.07$ \\
				\hline
				
			\end{tabular}
		\end{center}
	\end{table}   

\vspace{2em}

\section*{Acknowledgements}

DB, SM, and PS acknowledge the Indo-US Science and Technology Forum (IUSSTF) for supporting the current research. This research has been conducted using processed Wind data. We sincerely
thank the Wind support team. We further acknowledge the Wind (\url{https://wind.nasa.gov/mfi_swe_plot.php}) and OMNI (\url{https://omniweb.gsfc.nasa.gov/}) data sources. This paper has benefited from helpful inputs from two anonymous reviewers.

\vspace{-1em}


	\bibliographystyle{plain}
	\bibliography{jaa_plain.bib}

\end{document}